\documentclass[aip,jcp,amsmath,amssymb,reprint,superscriptaddress,longbibliography]{revtex4-1}

\usepackage{color}
\usepackage{graphicx}
\usepackage{txfonts}
\usepackage{mathrsfs}

\newcommand*\rev{\textcolor{black}}

\newcommand{\beginsupplement}{%
        \setcounter{table}{0}
        \renewcommand{\thetable}{S\arabic{table}}%
        \setcounter{figure}{0}
        \renewcommand{\thefigure}{S\arabic{figure}}%
        \setcounter{equation}{0}
        \renewcommand{\theequation}{S\arabic{equation}}%

        \setcounter{page}{1}
        \renewcommand{\thepage}{S\arabic{page}}%
}

\newcommand*\pb{$p_\mathrm{B}$}

\newcommand*\pbk{$p_{\mathrm{B}}^{(k)}$}
\newcommand*\fbqk{$f_\mathrm{B}(q^{(k)})$}
\newcommand*\xk{$\mathbf{x}^{(k)}$}
\newcommand*\qk{$q^{(k)}$}
\newcommand*\vele{V${}_{\mathrm{ele}}$(H${}_{18}$)}

\newcommand*\nlayer{N${}_{\mathrm{layer}}$}
\newcommand*\nnode{${\mathbf{N}}_{\mathrm{node}}$}
\newcommand*\lmd{$\lambda$}

\newcommand*\titlename{Investigating the hyperparameter space of deep neural network models for reaction coordinates}

\begin{document}

\title{\titlename}

\author{Kyohei Kawashima}
\affiliation{Institute for Materials Chemistry and Engineering, Kyushu University, Kasuga, Fukuoka 816-8580, Japan}

\author{Takumi Sato}
\affiliation{Department of Interdisciplinary Engineering Sciences, Interdisciplinary Graduate School of Engineering Sciences, Kyushu University, Kasuga, Fukuoka 816-8580, Japan}

\author{Kei-ichi Okazaki}
\email{keokazaki@ims.ac.jp}
\affiliation{Research Center for Computational Science, Institute for Molecular Science, Okazaki, Aichi 444-8585, Japan}
\affiliation{Graduate Institute for Advanced Studies, SOKENDAI, Okazaki,
Aichi 444-8585, Japan}

\author{Kang Kim}
\email{kk@cheng.es.osaka-u.ac.jp}
\affiliation{Division of Chemical Engineering, Department of Materials Engineering Science, Graduate School of Engineering Science, Osaka University, Toyonaka, Osaka 560-8531, Japan}

\author{Nobuyuki Matubayasi}
\email{nobuyuki@cheng.es.osaka-u.ac.jp}
\affiliation{Division of Chemical Engineering, Department of Materials Engineering Science, Graduate School of Engineering Science, Osaka University, Toyonaka, Osaka 560-8531, Japan}

\author{Toshifumi Mori}
\email{toshi{\_}mori@cm.kyushu-u.ac.jp}
\affiliation{Institute for Materials Chemistry and Engineering, Kyushu University, Kasuga, Fukuoka 816-8580, Japan}
\affiliation{Department of Interdisciplinary Engineering Sciences, Interdisciplinary Graduate School of Engineering Sciences, Kyushu University, Kasuga, Fukuoka 816-8580, Japan}

\date{\today}


\begin{abstract}
Identifying reaction coordinates (RCs) is a key to understand the mechanism of reactions in complex systems. Deep neural network (DNN) and machine learning approaches have become a powerful tool to find the RC. On the other hand, the hyperparameters that determine the DNN model structure can be highly flexible and are often selected intuitively and in a non-trivial and tedious manner. Furthermore, how the hyperparameter choice affects the RC quality remains obscure. Here, we explore the hyperparameter space by developing the hyperparameter tuning approach for the DNN model for RC, and investigate how the parameter set affects the RC quality. The DNN model is built to predict the committor along the RC from various collective variables by minimizing the cross-entropy function; the hyperparameters are automatically determined using the Bayesian optimization method. The approach is applied to study the isomerization of alanine dipeptide in vacuum and in water, and the features that characterize the RC are extracted using the explainable AI (XAI) tools. The results show that the DNN models with diverse structure can describe the RC with similar accuracy, and furthermore, the features analyzed by XAI are highly similar. This indicates that the hyperparameter space is multimodal. The electrostatic potential from the solvent to the hydrogen H${}_{18}$ plays an important role in the RC in water. The current study shows that the structure of the DNN models can be rather flexible while the suitably optimized model share the same features, thus a common mechanism from the RC can be extracted.
\end{abstract}


\maketitle 

%
%
\section{Introduction} \label{sec:introduction}

\begin{figure}
\includegraphics[width=6.8cm]{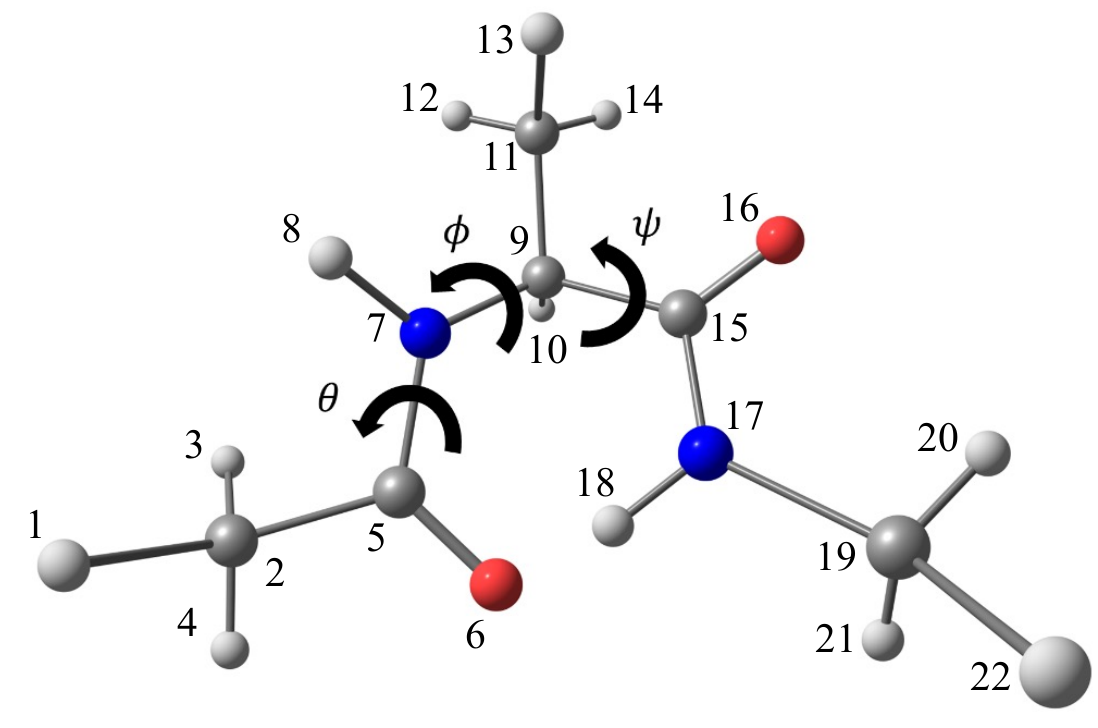}
\caption{Alanine dipeptide with atom indexes. The three key dihedral angles, i.e., $\phi$, $\psi$, and $\theta$, the dihedral angles about the C${}_5$-N${}_7$, N${}_7$-C${}_9$, and C${}_9$-C${}_{15}$ bonds, respectively, are also shown.}
\label{fig:molecule}
\end{figure}

Transition state (TS) plays a fundamental role in chemical reactions and enzyme catalysis.\cite{Pauling.1948,Wolfenden.1969,Schramm.2007,Jiang.2008,Peters.2017}
While reactions in simple model system can be well characterized using a few TSs, the reactions in solution and biomolecules involve numerous intermediate and TSs on the high-dimensional potential energy surface.\cite{Bruice.2002,Garcia-Viloca.2004,Ishida.2003,Hayashi.2012,Masgrau.2015}
It is thus indispensable to describe the reactions on the low-dimensional free energy surface (FES) described by a few collective variables (CVs), and it has been of great interest to determine the optimal CVs from a number of possible CV candidates that can adequately describe the FES and TS.
In a representative case of alanine dipeptide, the conformational transitions have been successfully characterized with the FES using two dihedral angles, $\phi$ and $\psi$, known as the Ramachandran plot\cite{Ramachandran.1968} (Figure \ref{fig:molecule}).

From a kinetic perspective, TS is considered as a point in potential or free energy surface having equal probability of reaching the reactant and product states.\cite{Geissler.1999,Hummer.2004,Peters.2016}
The committor $p_\mathrm{B}(\mathbf{x})$, defined as the probability of reaching the state $B$ from the configuration denoted by $\mathbf{x}$ without returning to state $A$, changes monotonically from 0 to 1 along an ideal RC and become 0.5 at the TS.
{\pb} thus serves as a good measure to evaluate the quality of a RC, and have been used to optimize the RCs from a large number of candidate CVs.\cite{Bolhuis.2000,Rhee.2005,Ma.2005,Peters.2007,Jung.2023,Quaytman.2007,Branduardi.2007,Mullen.2014,Okazaki.2019,Mori.2020,Mori.2020z1,Wu.20225z,Wu.20221kk,Manuchehrfar.2021,Zhang.2024}
These studies of RCs have led to realize that an adequate RC that satisfies the $p_\mathrm{B} \sim 0.5$ condition is often much more complicated than typical CVs used to build the FESs\cite{Ma.2005,Mullen.2014,Mori.2013}. For instance, in the case of alanine dipeptide, the $C_{\mathrm{eq}}$ to $C_{\mathrm{ax}}$ transition in vacuum required another dihedral angle $\theta$ (Figure \ref{fig:molecule})\cite{Bolhuis.2000}, and the reaction in water is even more complex and challenging.\cite{Bolhuis.2000,Ma.2005}
In particular, the RC in water have only been characterized successfully with an elaborated reaction coordinate, i.e., the solvent-derived electrostatic torque around one of the main-chain bonds, showing the challenge in identifying the RC coordinate when solvent is present.\cite{Ma.2005}

Machine learning (ML) approaches have recently been actively applied to identify the CVs and RCs from the molecular dynamics (MD) simulation trajectories\cite{Mardt.2018,Chen.201868,Sultan.2018,Ribeiro.2018,Rogal.2019,Wang.2020oz,Bonati.2020,Belkacemi.2021,Bonati.2021,Hooft.2021,Frassek.2021dp,Zhang.20213u,Kikutsuji.2022,Neumann.2022,Liang.2023,Lazzeri.2023,Singh.2023,Jung.2023,Ray.2023c38,Naleem.2023,Majumder.2024}.
In particular, Ma and Dinner\cite{Ma.2005} have developed an automatic CV search algorithm by combining the genetic algorithm to identify the RC for the alanine dipeptide isomerization.
Along this line, we have recently combined the deep neural network (DNN) approach with the cross-entropy minimization method\cite{Mori.2020,Mori.2020z1} to optimize the RC from many CV candidates using the committor distribution as a measure\cite{Kikutsuji.2022,Okada.2024}.
Our study highlighted the effectiveness of the explainable AI (XAI) method in characterizing the CVs that contribute to the RC.

While DNN approach can be very powerful in describing the non-linear contributions of CVs to RC, the structure of the DNN model characterized by the hyperparameters, e.g. the number of layers and nodes per layer, is highly flexible and are often chosen intuitively.
Yet, the adequacy of the hyperparameters remain ambiguous.
In this regard, Neumann and Schwierz\cite{Neumann.2022} have recently applied the Keras Tuner random search hyperparameter optimization to automatically determine the DNN model that can describe the committor of the magnesium binding to RNA.
Nevertheless, how the choice of hyperparameters affects the quality of the DNN model and the outcome remains unclear.
The applications of DNN models thus suffer from determining the appropriate hyperparameters, which remains to be a highly tedious and non-trivial task.

In this work, we developed a hyperparameter tuning protocol which utilizes the Bayesian optimization method with Gaussian process\cite{WU201926} to determine the DNN model for RC optimization.
The method takes the committors and large number of candidate CVs as the input without pre-determined hyperparameters for the DNN model, and automatically determines the appropriate DNN model for the RC.
The RC for the isomerization of alanine dipeptide in vacuum and in water are studied.
The diversity in the optimized hyperparameters and its effect to the RCs are explored, and the character of the RC for each model is analyzed by applying the XAI methods.
Furthermore, we show that the reaction in solution can be characterized using a more straightforward set of CVs while the complexity in the RC can be taken into account through the DNN framework.

%
%

%
%
\section{Methods} \label{sec:methods}
\subsection{Committor and Cross Entropy function}

Committor $p_\mathrm{B}(\mathbf{x})$ describes the probability of a trajectory generated from $\mathbf{x}$ to reach the product state $B$.
Along a RC $q$, {\pb} is expected to change smoothly from 0 to 1 and becomes 0.5 at the transition state.
The ideal committor value at $q$, $f_\mathrm{B}(q)$, can thus be described by a sigmoidal function, $f_\mathrm{B}(q) = \left( 1 + \tanh(q) \right)/2$.

When the data point $k$ is characterized by the collective variables (CVs) {\xk} and committor {\pbk}, the discrepancy between the ideal value and raw committor data can be described by the cross-entropy function\cite{Mori.2020,Mori.2020z1}
\begin{eqnarray}
\mathcal{H}(q) = - \sum_{k}{} p_\mathrm{B}^{(k)} \log f_\mathrm{B}(q^{(k)})
- \sum_{k}{} \left( 1 - p_\mathrm{B}^{(k)}\right) \log\left\{ 1 - f_\mathrm{B}(q^{(k)})\right\} \label{eq:cross_entropy}
\end{eqnarray}
Here, $k$ denotes the data points and {\qk} is the RC as a function of {\xk}.
Eq. \ref{eq:cross_entropy} is derived from the Kullback-Leibler divergence\cite{Mori.2020}, which quantifies the mismatch between the distribution of the raw ({\pbk}) and expected ($f_\mathrm{B}(q^{(k)})$) committors.
It is also noted that Eq. \ref{eq:cross_entropy} is a generalization of the maximum-likelihood function used with aimless shooting algorithm\cite{Peters.2007}.

By minimizing Eq. \ref{eq:cross_entropy}, one can optimize {\qk} to minimize the difference between {\pb} and {\fbqk}.
In practice, we employ the $L_2$ regularization function to suppress overfitting, and thus the loss function is defined as a sum of $\mathcal{H}$ and the penalty term\cite{Mori.2020}.
The regularization parameter {\lmd} is set separately for each layer (see below).

\subsection{Deep neural network model}

\begin{figure}
\includegraphics[width=8.5cm]{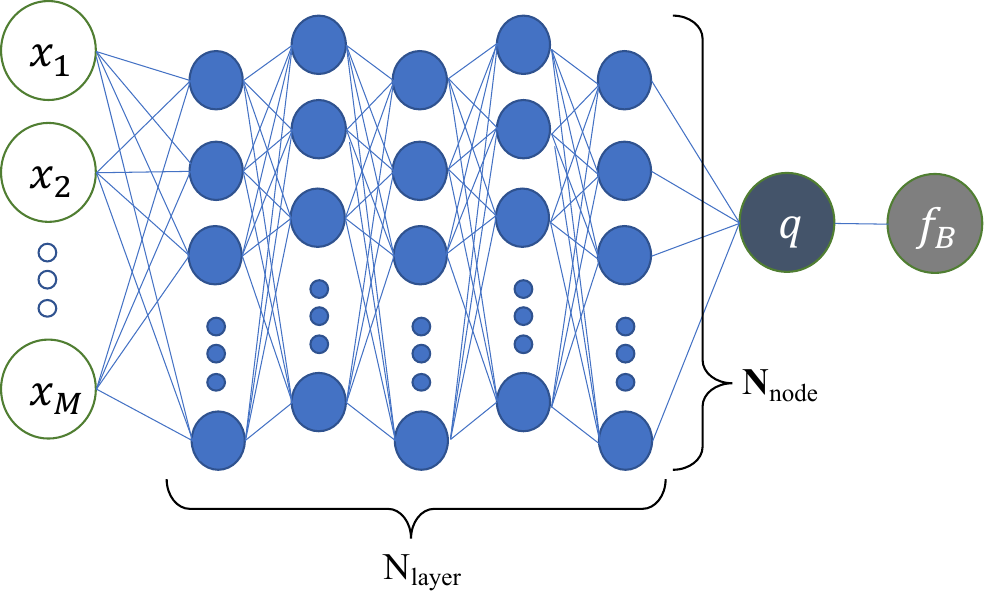}
\caption{Overview of the multilayer perception model used in this work. The input CVs ($x_i$) are converted to the output ($q$) with a DNN model of {\nlayer} consisting of {\nnode}. $f_\mathrm{B}$ is the predicted {\pb} value at $q$ described by a sigomidal function ($f_\mathrm{B}(q) = (1 + \tanh(q))/2$).}
\label{fig:image_dnn}
\end{figure}

The DNN function converts the CVs {\xk} into a RC {\qk} in a non-linear manner.
Here we adopt the multilayer perceptron model which consists of the input layer {\xk}, multiple hidden layers with different number of nodes, and the output layer {\qk} (Figure \ref{fig:image_dnn}).
The CVs are standardized prior to constructing {\xk}.
The leaky rectified linear unit  (Leaky ReLU) with a leaky parameter set of 0.01 was used for the activation function.
The $L_2$ regularization was employed, where {\lmd} was varied in the hyperparameter tuning step.
Note that different {\lmd} was used for each layer.
The numbers of hidden layers and nodes in each layer, {\nlayer} and {\nnode}, respectively, are also the hyperparameters that are left to be explored.
Optimization  was performed using AdaMax. The learning rate  {\it lr} and two decay  factors $\beta_1$ and $\beta_2$ were set to the default values of 0.001, 0.9, and 0.99, respectively.
The TensorFlow\cite{tensorflow2015-whitepaper} library was used to implement the DNN model.

%
%

%
%
\subsection{Hyperparameter tuning and DNN model optimization} \label{subsec:hyperparameter}

The hyperparameters in the current DNN, {\nlayer}, {\nnode}, and {\lmd}, are not unique and can strongly affect the performance of the DNN model.
Here we employ the hyperparameter tuning approach using the Bayesian optimization method with Gaussian process\cite{WU201926} to determine these parameters in an automatic manner.

The hyperparameter tuning and DNN model optimization are performed in two stages.
First, the data is divided into training and validation set at a ratio of 8:2.
{\nlayer} is searched between 2 and 5, and {\nnode} are explored separately for each layer, which are chosen within the range from 100 to 5000 in 100 increments.
{\lmd} is explored between 0.0001 to 0.100 with 20 points equally spaced in logarithmic scale.
The initial values for these parameters are chosen randomly within the exploration range.
Bayesian optimization was performed for 150 trials unless otherwise noted, where the DNN model for each hyperparameter was trained for a maximum of 1000 epochs using Eq. \ref{eq:cross_entropy}.
Early stopping was applied when the value of loss function for the validation set did not improve for 5 consecutive steps.
The performance of the model at each step was evaluated using the root-mean-square error (RMSE) for the validation set, calculated as a difference between the predicted and reference $p_\mathrm{B}$.
After the optimal choice of the parameters are determined, the data are unified and re-partitioned into training, validation, and test sets at a ratio of 5:1:4, and the DNN model is optimized for 1000 epochs with the same criteria for Early Stopping.
The Keras Tuner\cite{omalley2019kerastuner} library was used to implement the hyperparameter tuning with Gaussian process.

%
%
\subsection{DNN model interpretation with LIME and SHAP} \label{subsec:xai}

To interpret the DNN model, the Local Interpretable Model-agnostic Explanation (LIME)\cite{10.1145/2939672} and SHapley Additive exPlanations (SHAP)\cite{NIPS2017_8a20a862} methods were applied to the data.
In brief, LIME builds a linear regression function to explain the local behavior of the target data from the perturbation of input variables, whereas SHAP employs an additive feature attribution method that ensures fair distribution of predictions among input features in accordance with the game-theory-based Shapley value. 
LimeTabularExplainer for LIME and DeepExplainer for SHAP were employed, using the packages obtained from https://github.com/marcotcr/lime and https://github.com/slundberg/shap, respectively.

%
%
\subsection{Conformational sampling of alanine dipeptide} \label{subsec:md}
The initial structures of alanine dipeptide in vacuum and water were generated using AmberTools21\cite{Case.2021}.
In the case in water, alanine dipeptide was solvated in a rectangular parallelepiped box with 1683 water molecules.
Alanine dipeptide and water were treated with the Amber14SB force field and TIP3P model, respectively.
The system was energy minimized for 3000 steps and heated up to 300 K in 50 ps. MD simulations under NPT and NVT conditions were then performed for 100 and 1000 ps, respectively, to complete equilibration.

Umbrella sampling (US) and transition path sampling (TPS) were combined to collect broad conformations about the transition path of alanine dipeptide in vacuum and water as follows.
First, the transition state region (i.e., $\phi \sim$ 0) was sampled using the US along $\phi$ with a force constant of 100.0 kcal mol${}^{-1}$ rad${}^{-2}$ while restraining $\psi$ to $\psi \le 0^{\circ}$ with a half-harmonic potential with a force constant of 10.0 kcal mol${}^{-1}$ rad${}^{-2}$ (See Figure \ref{fig:molecule} for definitions of $\phi$ and $\psi$).
100 conformations from the 10 ns-long trajectory with the harmonic restraint centered at $\phi = 0.0$ was then evenly collected.
From each conformation, 10 trajectories of 2 ps long were generated by sampling the initial velocity from the Maxwell-Boltzmann distribution at 300 K and propagating the trajectory for 1 ps forward and backward in time under the NVE condition.
States A and B were defined as $\phi \le -30$ and $\phi \ge 30$, respectively.
It is noted that while the previous studies\cite{Ma.2005,Bolhuis.2000} have often studied the C${}_{\mathrm{7eq}}$ $\rightarrow$ $\alpha_{\mathrm{R}}$ transitions in solution, which have lower energy barrier\cite{Anderson.1988}, here we focus on the isomerization about $\phi$ to compare the results in vacuum and water.

From the successful transition trajectories which connect states A and B, new points were generated by extracting snapshots within 0.1 ps from the time origin of each trajectory.
The velocity for each point was sampled from the Maxwell-Boltzmann distribution at 300 K, and a new ensemble of 2 ps-long trajectories were generated following the same procedure as above.
Roughly 33 and 42 {\%} of the trial trajectories were accepted throughout the TPS of alanine dipeptide in vacuum and water, respectively.
After the 3rd round, 3,714 and 4,590 points were generated for the following committor calculations in vacuum and in water.

\begin{figure}
\includegraphics[width=8.5cm]{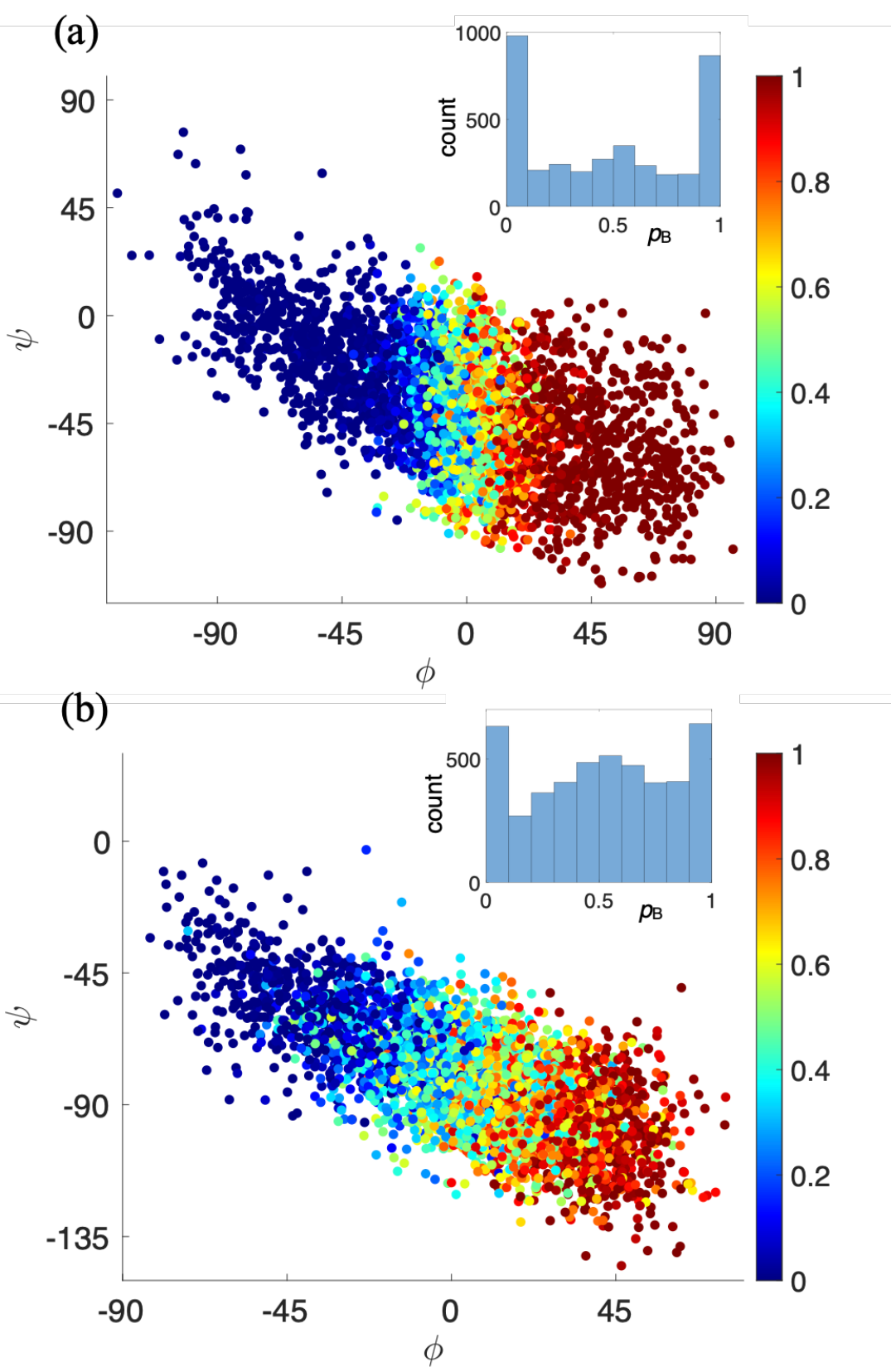}
\caption{Distributions of the data points in ($\phi$,$\psi$) space from the trajectories in (a) vacuum and (b) water. Colors represent the calculated $p_\mathrm{B}$ data at each point. The insets in (a) and (b) show the committor distributions for all the data points in vacuum and water, respectively.}
\label{fig:phipsi_hist}
\end{figure}

From each data point after the final round of TPS, 2 ps-long trajectories were generated 100 times per point with velocities randomly sampled from the Maxwell-Boltzmann distribution at 300 K.
$p_\mathrm{B}$ was calculated by $p_\mathrm{B} = n_{B} / ( n_{A} + n_{B} ) $ where $n_I$ denotes the number of trajectories that is at state $I$ at the end of the trajectory.
The data points projected on the ($\phi$, $\psi$) plane, with colors describing the $p_\mathrm{B}$ value, are shown in Figure \ref{fig:phipsi_hist}.
We note that the data points were obtained over a broader range of $\phi$ compared to those sampled with the aimless shooting algorithm\cite{Mori.2020z1}.
As a consequence, the $p_\mathrm{B}$-distribution were not even, especially in vacuum where roughly half of the points were found in either $p_\mathrm{B} \le 0.1$ or $p_\mathrm{B} \ge 0.9$.
Here  we note that the minima on the $\phi < 0$ side in water, corresponding to the $\alpha_{\mathrm{R}}$ state, is located at $\psi < 0$\cite{Ma.2005,Anderson.1988}.
The transition path in water is thus also located on the $\psi < 0$ side compared to that in vacuum.

The collective variables (CVs) are listed in Supplementary material Table \ref{table:si_cvs}, and the atom indexes are given in Figure \ref{fig:molecule}.
All MD simulations were performed using Amber 20 software package\cite{Case.2021,Salomon-Ferrer.2013}.

%
%

%
%
\section{Results and discussion} \label{sec:results}
%
%
\subsection{Convergence of optimal hyperparameters} 
\begin{figure}
\includegraphics[width=8cm]{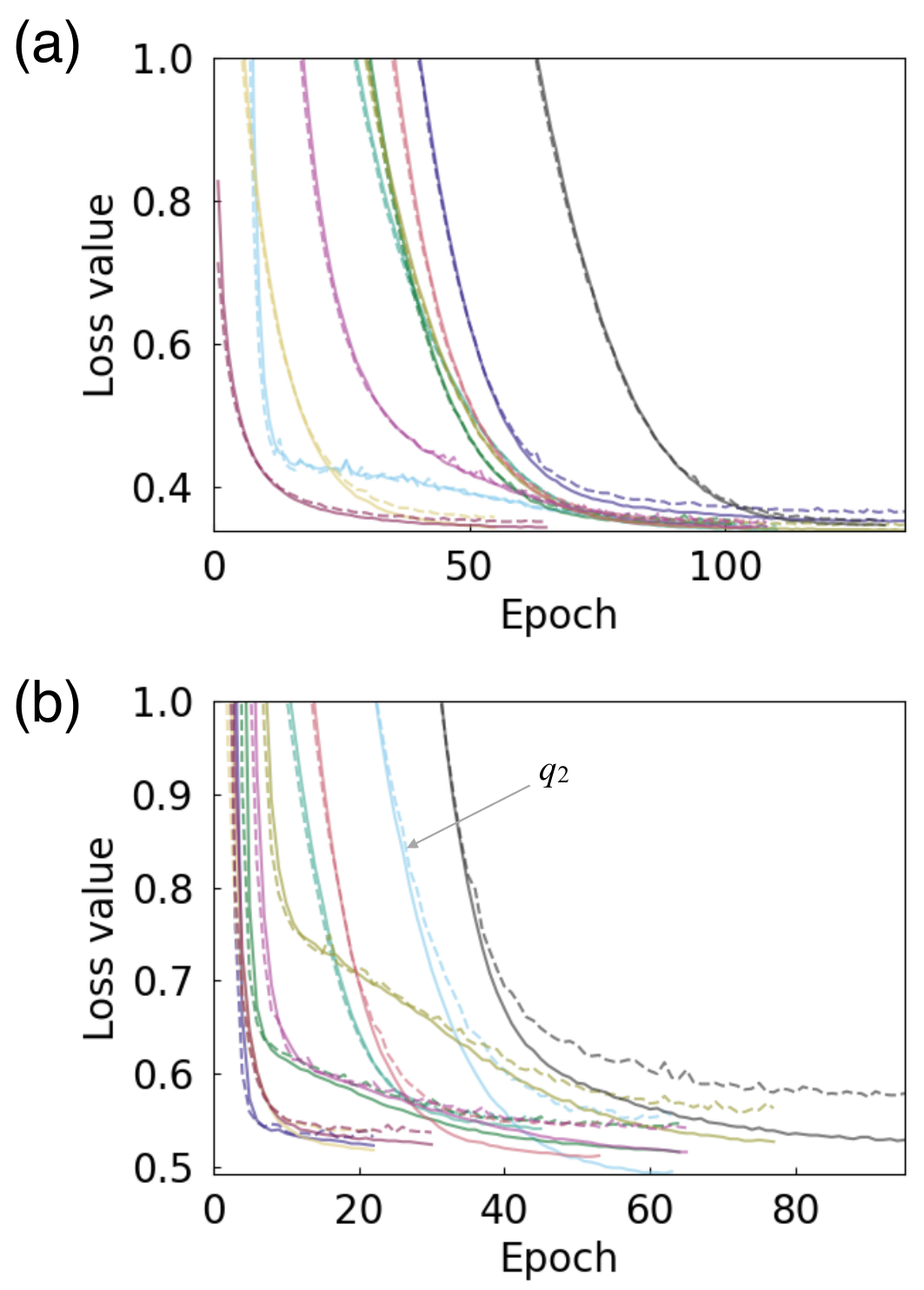}
\caption{Training process of the loss function as a sum of the cross-entropy function and the $L_2$ regularization term in (a) vacuum and (b) water.
Full and dashed lines show the values for training and validation data along the training process, and colors denote results for different models.
Note that the training and data sets in each model are different due to different initial seed for data partitioning.
}
\label{fig:learn_loss}
\end{figure}

We first compare the results of hyperparameter tuning for the isomerization  of alanine dipeptide in vacuum.
10 models were constructed using different initial seeds including data partitioning.
The CV candidates consist of 45 dihedral angles in cosine and sine forms, i.e., 90 variables, which follow our previous works.\cite{Mori.2020z1,Kikutsuji.2022}
The convergence of the loss function, given in Fig.\ref{fig:learn_loss}, shows that the optimizations are converged within 60 to 100 epochs.
Since further extending the number of epochs leads to slightly decrease the loss function for the training data but increase in that for the validation data, we find that these epoch numbers are sufficient for suppressing overfitting and maximizing predictability.
The obtained hyperparameters are summarized in Table \ref{table:opt_hyperparam_vac}.
Surprisingly, the number of layers ({\nlayer}) and nodes ({\nnode}) differ remarkably between the models.
{\nnode} often converged to the maximum (5000) or to the minimum (100), and {\nnode} = 5 was most frequently selected (5 out of 10 models).
{\lmd} was often found to be at the minimum (0.0001), especially in the first layer, but can be up to 0.0695 and varied between the layers.
We note that before converging to these values, different regions of the hyperparameter space have been explored before converging to the optimal values in each model (Fig. \ref{fig:si_hp_history_vac}). 
Overall, no apparent unique optimal model was obtained.

%
%
\begin{table*}
\caption{Optimized hyperparameters in vacuum. $q_i$ denote the RC obtained for the $i$-th model.}
\centering
\begin{tabular}{l|rrrrrrrrrr}
    \hline \hline
     & $q_1$ &   $q_2$ &   $q_3$ &   $q_4$ &   $q_5$ &   $q_6$ & $q_7$ & $q_8$ & $q_9$ & $q_{10}$
    \\
{\nlayer} &5 &2 &3 &5 &5 &3 &5 &4 &2 &5 \\
\hline
$\mathbf{N}_{\mathrm{node}}^{(1)}$ &5000 &4000 &4000 &3500 &5000 &1400 &5000 &4100 &2700 &5000 \\
{\lmd}${}^{(1)}$ & 0.0001 & 0.0001 & 0.0001 & 0.1000 & 0.0001 & 0.0001 & 0.0001 & 0.0009 & 0.0001 & 0.0001 \\
$\mathbf{N}_{\mathrm{node}}^{(2)}$ &5000 &5000 &2500 &5000 &5000 &5000 &5000 &2700 &100 &2100 \\
{\lmd}${}^{(2)}$ & 0.0001 & 0.0113 & 0.0026 & 0.0001 & 0.1000 & 0.0001 & 0.0013 & 0.0018 & 0.0009 & 0.1000  \\
$\mathbf{N}_{\mathrm{node}}^{(3)}$ &5000 &--&2900 &100 &4300 &100 &5000 &3600 &  --&5000 \\
{\lmd}${}^{(3)}$ &0.0695 &--&0.0001 &0.0001 &0.0001 &0.0001 &0.0001 &0.0006  &--&0.0026 \\
$\mathbf{N}_{\mathrm{node}}^{(4)}$ &3600 &--&--&2400 &100 &  --&5000 &900 &--&100 \\
{\lmd}${}^{(4)}$ & 0.1000 & -- & -- & 0.0002 & 0.0001 & -- & 0.1000 & 0.0002 & -- & 0.0001 \\
$\mathbf{N}_{\mathrm{node}}^{(5)}$ &100 &--&--&5000 &5000 &  --&2000 &--&--&5000 \\
{\lmd}${}^{(5)}$ &0.0001 &--&--&0.0001 &0.0001 &--&0.0001 &--&--& 0.0001
\\ \hline \hline
\end{tabular}
\label{table:opt_hyperparam_vac}
\end{table*}

\begin{figure}
\includegraphics[width=8.5cm]{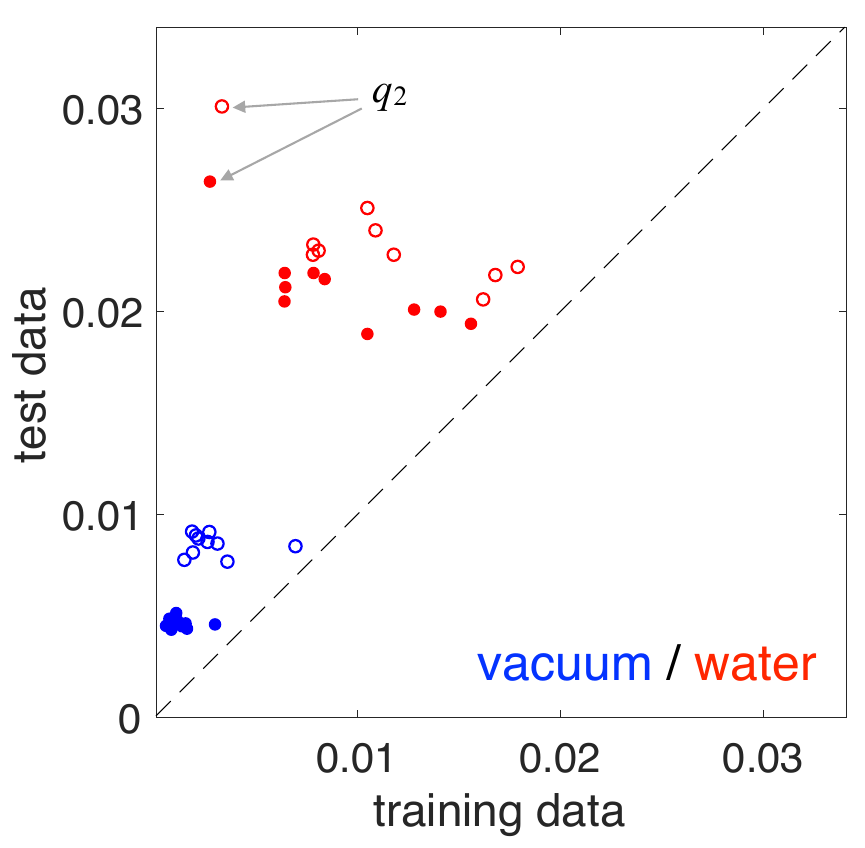}
\caption{Scatter plot for RMSEs between the predicted and reference {\pb} for the training and test data sets. Filled and open circles indicate the RMSEs using full points and those at about the TS ($-0.2 < q_i < 0.2$), respectively.
Blue and red colors are the results in vacuum and water.}
\label{fig:rmse_train_test}
\end{figure}

To compare the accuracy of the coordinates from the perspective of $p_\mathrm{B}$-predictability, the RMSE between the predicted and reference {\pb} for the training and test data are shown in Fig. \ref{fig:rmse_train_test} for the 10 models.
The RMSEs were within 0.005 and around 0.005 for the training and test sets, respectively.
Even the RMSEs for the data points about the TS ($-0.2 < q_i < 0.2$) are within 0.007 and 0.009 for the training and test data.
These results indicate that while the optimized DNN models are not unique, the RCs show very similar quality, i.e., able to predict {\pb} with similar accuracy.
This implies that the hyperparameter space for the current DNN model is likely multimodal.

\begin{figure}
\includegraphics[width=8.5cm]{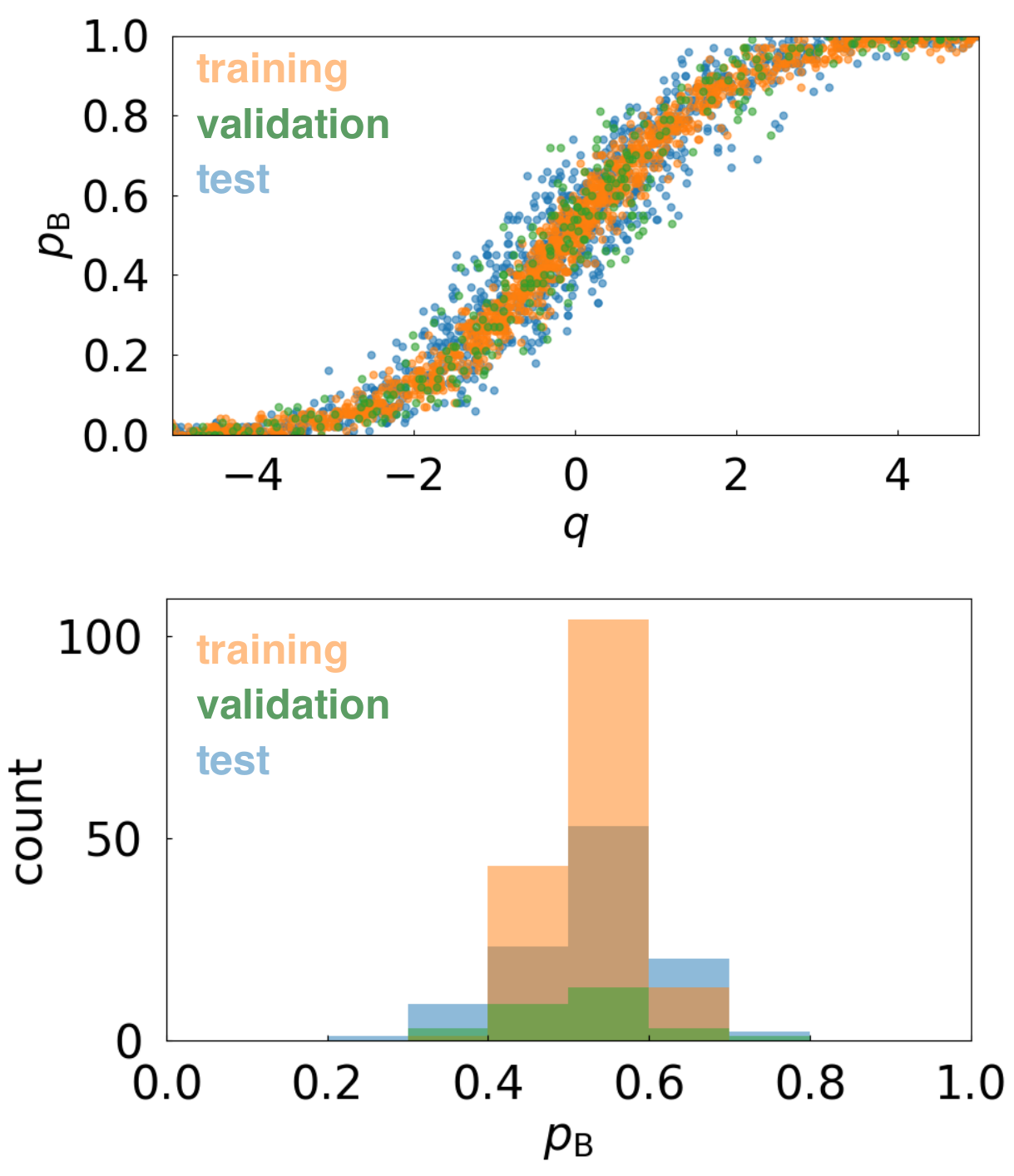}
\caption{(a) Scatter plots of the optimized coordinate ($q$) and committor ($p_\mathrm{B}$) for $q_1$ in vacuum. (b) Distribution of $p_\mathrm{B}$ for the points within $-0.2 < q_1 < 0.2$. Orange, green, and blue in (a) and (b) denote the results from the training, validation, and test data sets, respectively.}
\label{fig:sigmoid_ala_vac}
\end{figure}

Figure \ref{fig:sigmoid_ala_vac} summarizes the change of $p_\mathrm{B}$ along the first RC ($q_1$).
Figure \ref{fig:sigmoid_ala_vac}(a) shows that both training and test data closely follows the ideal sigmoidal line.
The histogram of $p_\mathrm{B}$ about the transition state (Fig. \ref{fig:sigmoid_ala_vac}(b)) indicates that both training and test data set shows a sharp peak at $p_\mathrm{B} \sim 0.5$.
These result implies that $q_1$ serves as a good RC and can clearly characterize the transition state.
We note that similar results are obtained for the other RCs (Figs. \ref{fig:si_sigmoid_ala_vac} and \ref{fig:si_hist_ala_vac}).

\begin{figure}
\includegraphics[width=8.5cm]{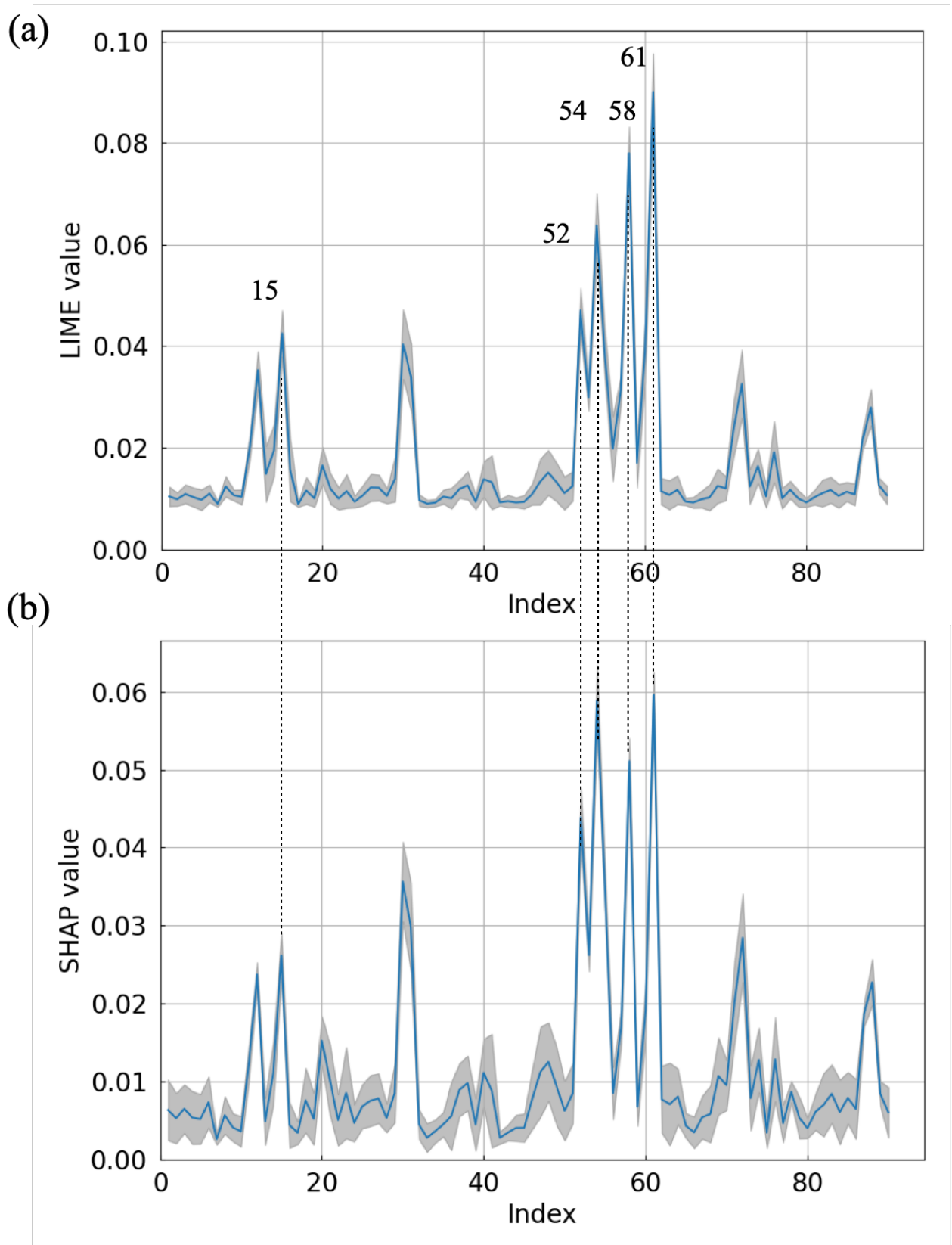}
\caption{Contributions of CVs to the RCs in vacuum extracted using (a) LIME and (b) SHAP in absolute values. Blue lines and gray shades denote the average and variance calculated from the 10 RCs.}
\label{fig:xai_vac}
\end{figure}

Figure \ref{fig:xai_vac} summarizes the features that contribute to RC at about the TS extracted from LIME and SHAP.
The extracted features are found to be very similar between the 10 RCs.
CV${}_{61}$, CV${}_{58}$, and CV${}_{54}$, corresponding to the sine form of $\phi$, $\phi$, and $\theta$, respectively, are the three major CVs in LIME analysis.
The contribution of $\theta$ is increased in the SHAP analysis, but key CVs are unchanged from those found in LIME.
The result is consistent with the previous studies which showed that $\theta$ becomes important at about the transition state\cite{Ma.2005,Kikutsuji.2022,Mori.2020z1}.
On the other hand, the order as well as magnitude of the contributions are slightly different from those obtained in the previous XAI analysis\cite{Kikutsuji.2022}.
This may be due to the differences in the distribution of the data points where the current points are distributed over a broader range along $\phi$ (Fig. \ref{fig:phipsi_hist}).

We also directly compared the RCs obtained from different DNN models in the scatter plots and correlation coefficients (Fig. \ref{fig:si_corr_vac}).
The result shows that every pair of RCs is highly correlated and the correlation coefficient is $>$0.99, indicating that despite the difference in the hyperparameters, the obtained RC are very similar.
It is noted that the points at the negative and positive ends of the RCs where {\pb} is 0 and 1, respectively, show slight deviations from the diagonal line in some of the plots.
This is because {\pb} is insensitive to the changes in the RCs at these ranges, and thus these points are not further optimized.
The current results show that while the optimization of DNN hyperparameters do not lead to a unique model, each DNN model produces equally accurate RCs with similar predictability of {\pb} and common features that describe the TS.

%
%

%
%

%
%

%
%

%
%
\subsection{Characterization of solvation coordinates using DNN}

Next we apply the current approach to the isomerization of alanine dipeptide in water.
Compared to the case in vacuum, the reaction in water has been a challenging task due to the contribution from the numerous waters surrounding the solute.\cite{Bolhuis.2000}
Previously, a complex CV involving the solvent-derived electrostatic torque have been proposed to be important for the $\mathrm{C}_{\mathrm{7eq}} \rightarrow \alpha_{\mathrm{R}}$ transition\cite{Ma.2005}.
Here, we attempt to adopt more intuitive CVs to describe the solvent contribution and account for the complex non-linear contribution through the DNN model.
To this end, the electrostatic and van der Waals interaction from the waters to the atoms in alanine dipeptide were used as the CVs in addition to the internal coordinates of alanine dipeptide (i.e., dihedral angles).
Optimizations of the hyperparameters and DNN models were performed in the same manner as those in vacuum.

As is the case in vacuum, the optimized DNN models for the RCs in water were found to converged to different parameter sets (Table \ref{table:opt_hyperparam_wat}).
{\nlayer} = 3 most frequently appeared, and {\nnode} were on average slightly smaller than those in vacuum.
For instance, in the case of $q_1$, the hyperparameter space was found explored quickly before converging to the optimal value (Fig. \ref{fig:si_hp_history_wat}); the larger number of layers were explored but only for a short period in $q_1$; in the latter exploration stage, the {\nlayer} = 3 was most intensely searched, and {\nlayer} of 3 was found to be optimal.
On the other hand, {\lmd} were found to become larger than those in vacuum.
These trends indicate that the increase in the number of CVs in water resulted in a slightly more compact model but with higher regularization penalty to suppress over-fitting.

%
%
\begin{table*}
\caption{Optimized hyperparameters in water. $q_i$ denote the RC obtained for the $i$-th model.}
\centering
\begin{tabular}{l|rrrrrrrrrr}
    \hline \hline
     & $q_1$ &   $q_2$ &   $q_3$ &   $q_4$ &   $q_5$ &   $q_6$ & $q_7$ & $q_8$ & $q_9$ & $q_{10}$
    \\
{\nlayer} &3&5&3&3&5&2&3&2&3&5 \\
\hline
$\mathbf{N}_{\mathrm{node}}^{(1)}$ &100&4400&3800&2900&5000&1700&1300&2900&2300&5000 \\
$\lambda^{(1)}$ &0.0001 &0.0004 &0.0013 &0.0001 &0.0001 &0.0026 &0.0009 &0.0018 &0.0001 &0.0001 \\
$\mathbf{N}_{\mathrm{node}}^{(2)}$ &1200&1700&1600&5000&5000&100&800&100&5000&5000 \\
$\lambda^{(2)}$ &0.1000 &0.0162 &0.0483 &0.0079 &0.0079 &0.0018 &0.1000 &0.0234 &0.0483 &0.0055 \\
$\mathbf{N}_{\mathrm{node}}^{(3)}$ &3100&3800&1400&100&2000&  --&3000&  --&400&2500 \\
$\lambda^{(3)}$ &0.0055 &0.0004 &0.0336 &0.1000 &0.1000 &  --&0.0018 &  --&0.0234 &0.1000 \\
$\mathbf{N}_{\mathrm{node}}^{(4)}$ &  --&800&  --&  --&5000&  --&  --&  --&  --&2000 \\
$\lambda^{(4)}$ &  --&0.0003 &  --&  --&0.1000 &  --&  --&  --&  --&0.1000 \\
$\mathbf{N}_{\mathrm{node}}^{(5)}$ &  --&600&  --&  --&100&  --&  --&  --&  --&3700 \\
$\lambda^{(5)}$ &  --&0.0009 &  --&  --&0.0001 &  --&  --&  --&  --&0.0055 \\
\\ \hline \hline
\end{tabular}
\label{table:opt_hyperparam_wat}
\end{table*}

The RMSEs between the calculated and reference {\pb}s, plotted in Fig. \ref{fig:rmse_train_test}, show that the RMSEs for the training data are mostly distributed between 0.006 and 0.015 whereas those for the test data are found at around 0.02.
Only in one case we find slight sign of overfitting where the RMSE for the training and test data are 0.002 and 0.026, respectively.
The RMSEs for the data about the transition state ($-0.2 < q_i < 0.2$) show similar trend.
These result indicate that while the RMSEs for the data in water are larger than that in vacuum for both the training and test data, the optimized RCs can satisfactorily predict {\pb}s.

\begin{figure}
\includegraphics[width=8.5cm]{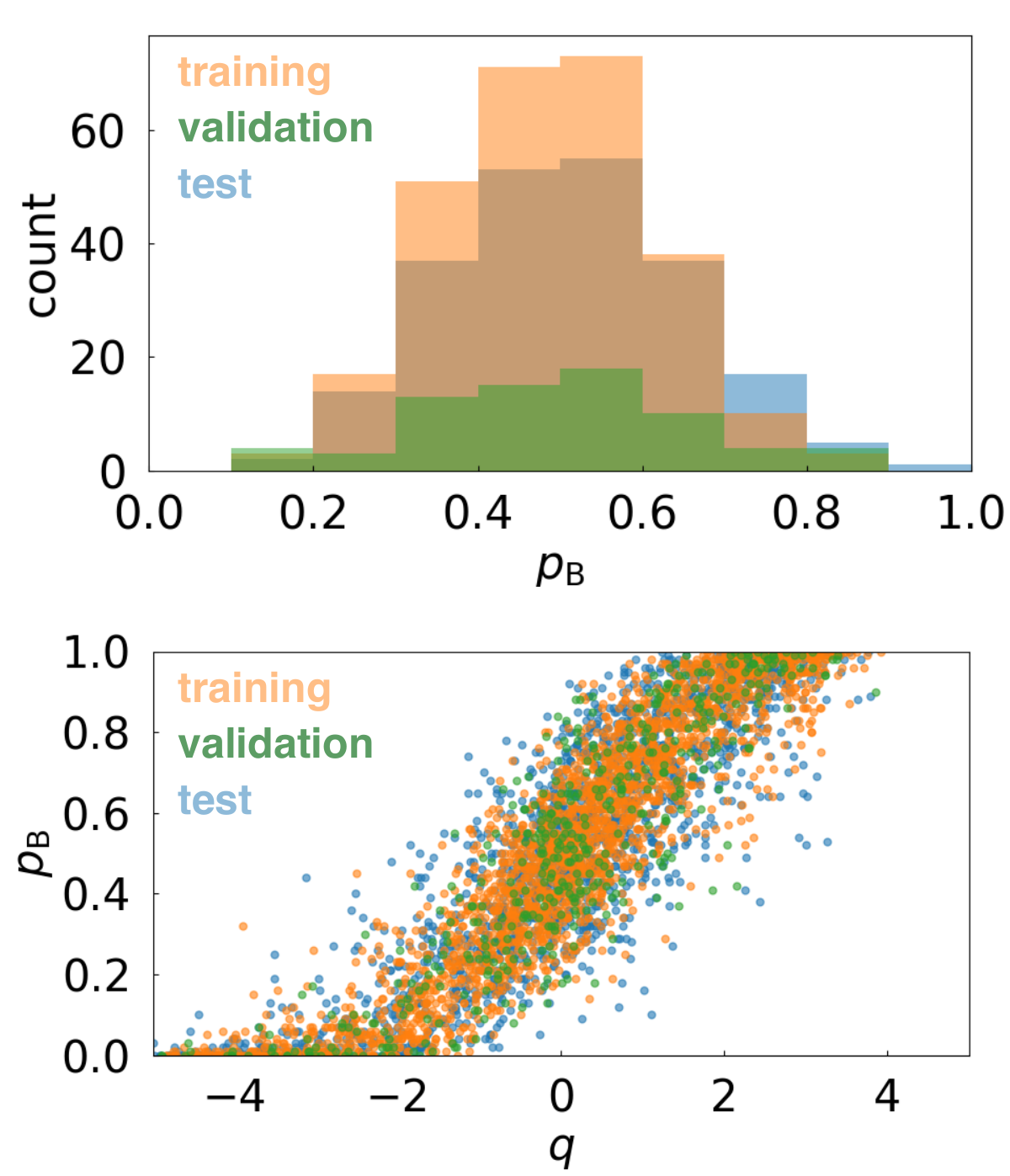}
\caption{(a) Scatter plots of the optimized coordinate ($q$) and committor ($p_\mathrm{B}$) for $q_1$ in water. (b) Distribution of $p_\mathrm{B}$ for the points within $-0.2 < q_1 < 0.2$. Orange, green, and blue in (a) and (b) denote the results from the training, validation, and test data sets, respectively.}
\label{fig:sigmoid_ala_wat}
\end{figure}

\begin{figure*}
\includegraphics[width=14cm]{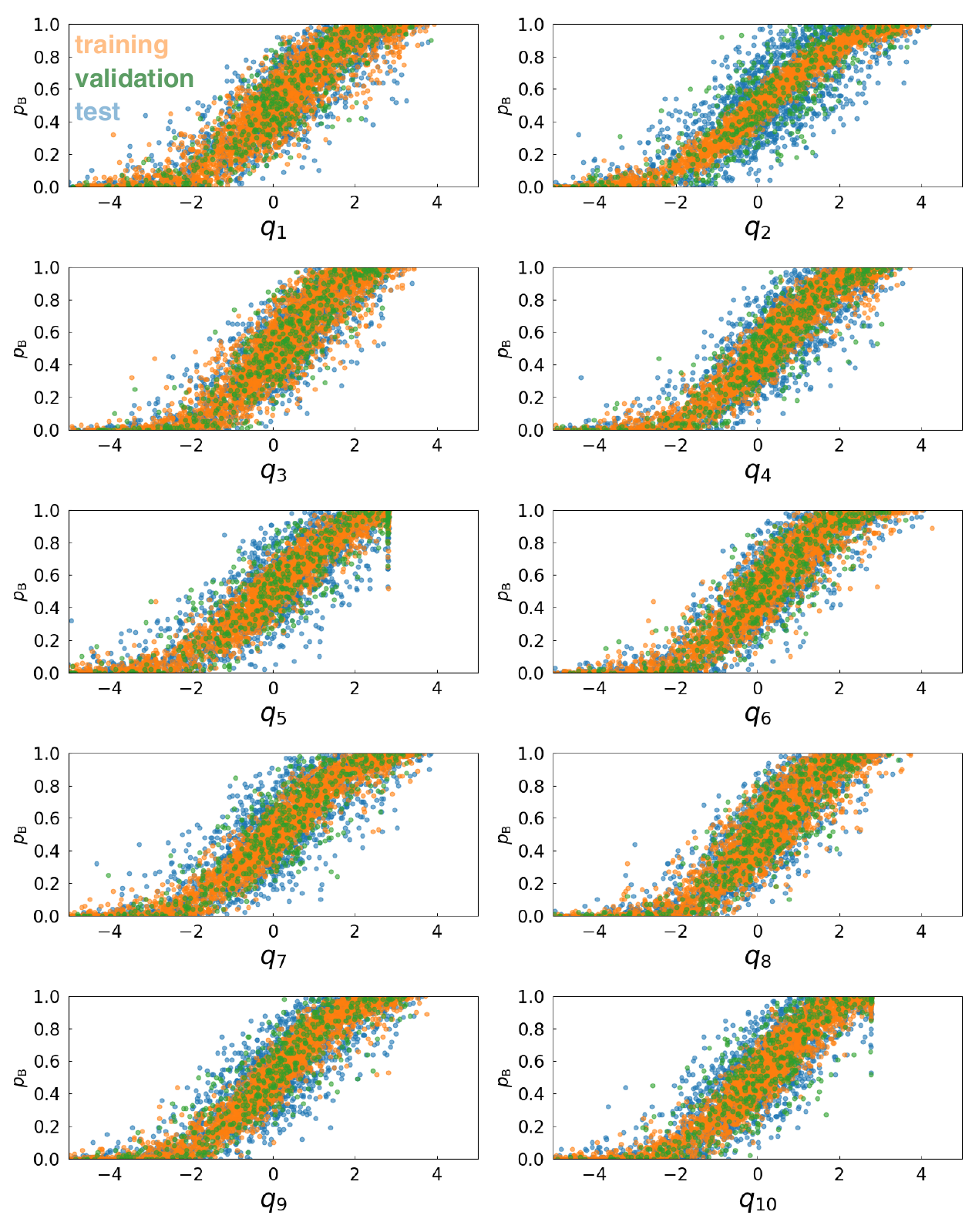}
\caption{Scatter plots of the optimized coordinates ($q_1$ to $q_{10}$) and committors ($p_\mathrm{B}$) in water. Orange, green, and blue in (a) and (b) denote the results from the training, validation, and test data sets, respectively.}
\label{fig:si_sigmoid_ala_wat}
\end{figure*}

The change of {\pb} along $q_1$ in water is summarized in Fig. \ref{fig:sigmoid_ala_wat}.
The results for other RCs are illustrated in Figs. \ref{fig:si_sigmoid_ala_wat} and  \ref{fig:si_hist_ala_wat}.
The distribution of {\pb} (Fig. \ref{fig:sigmoid_ala_wat}(a)) is broader than the case in vacuum but follows a sigmoidal curve as a function of $q_1$.
The histogram of {\pb} near the TS (Fig. \ref{fig:sigmoid_ala_wat}(b)) shows a peak at {\pb}$\sim$0.5 with the width of the histograms broader than those in vacuum.
The results for the other optimized coordinates in Fig. \ref{fig:si_sigmoid_ala_wat} are mostly consistent with $q_1$.
We note that only in the case of $q_2$, the histogram for the training data is sharply peaked while that for the test data is broad, implying that overfitting to the training data has occurred.
This is consistent with the loss function values and RMSE results (Figs. \ref{fig:learn_loss}(b) and \ref{fig:rmse_train_test}(b), respectively).

\begin{figure}
\includegraphics[width=8.5cm]{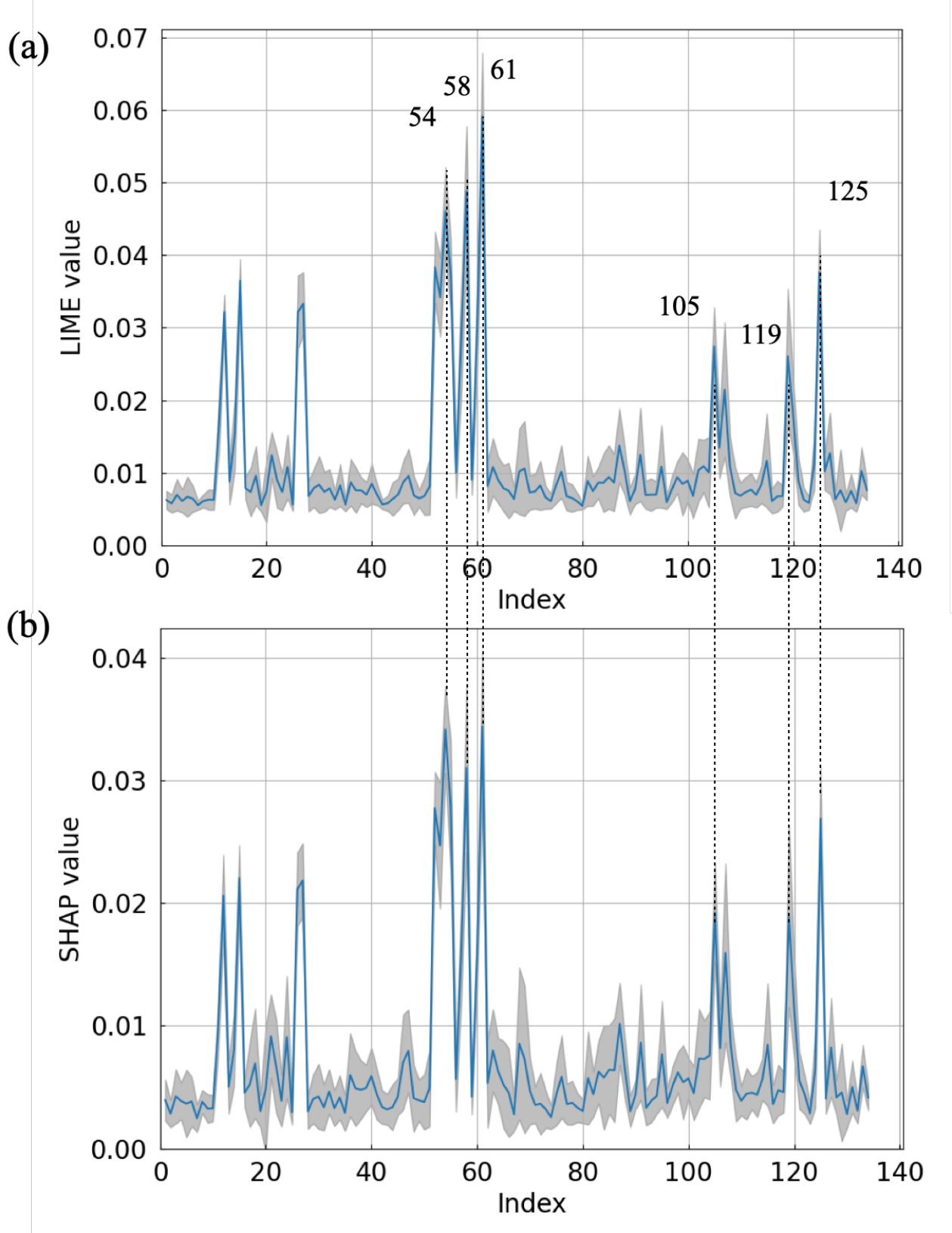}
\caption{Contributions of CVs to the RCs in water extracted using (a) LIME and (b) SHAP in absolute values. Blue lines and gray shades denote the average and variance calculated from the 10 RCs.}
\label{fig:xai_wat}
\end{figure}

\begin{figure*}
\includegraphics[width=17cm]{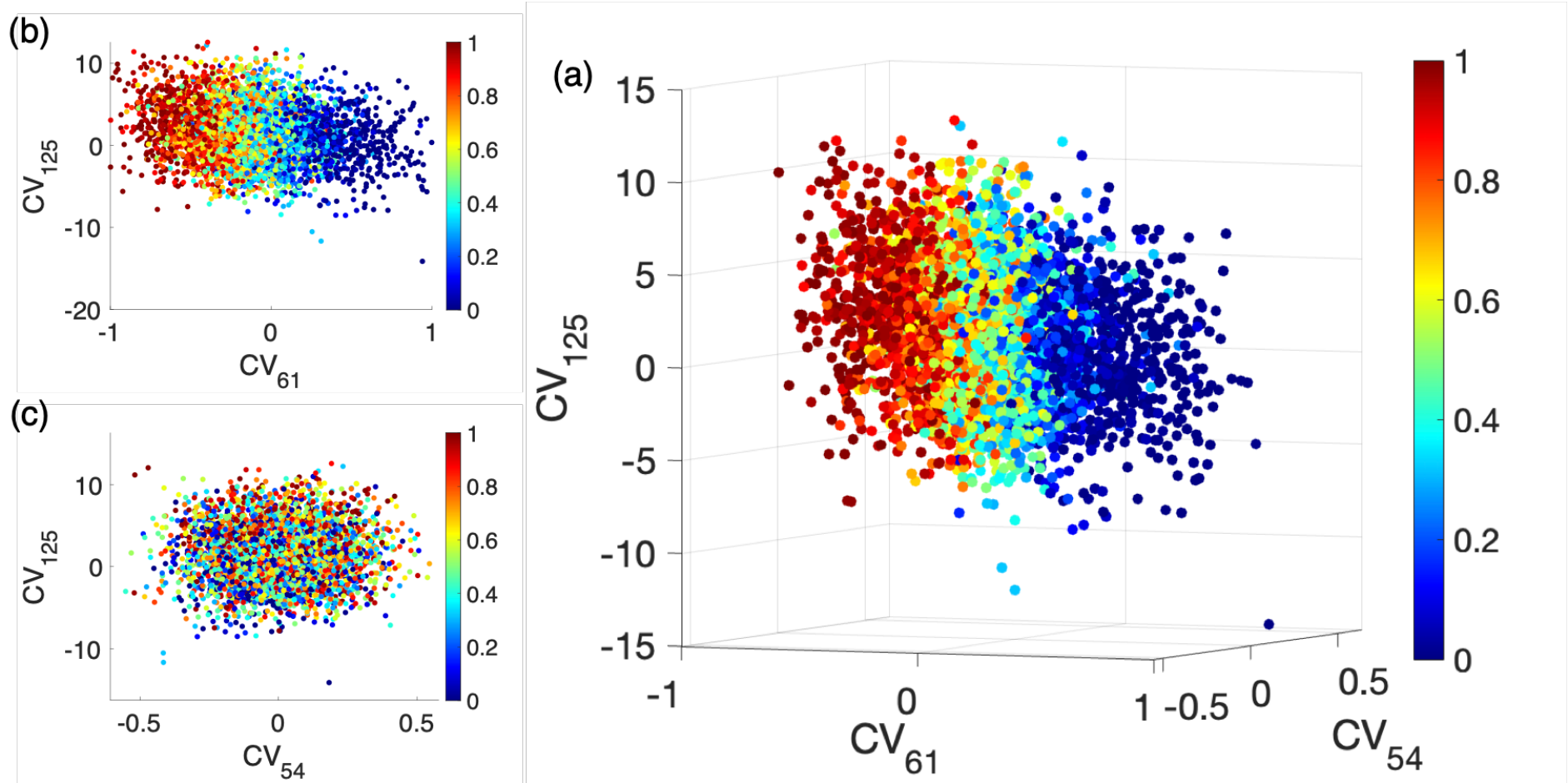}
\caption{Correlations between the changes in CVs and {\pb} in water from the scatter plots in the spaces of (a) CV${}_{61}$ ($\phi$) and CV${}_{125}$ ({\vele}), (b) CV${}_{56}$ ($\theta$) and CV${}_{125}$, and (c) CV${}_{61}$, CV${}_{56}$, and CV${}_{125}$. Color range denotes the calculated {\pb} at each point.}
\label{fig:corr_cvs_pb}
\end{figure*}

Figure \ref{fig:xai_wat} shows the contributions of CVs to the RCs at about the TS extracted by LIME and SHAP.
Note that these analyses include $q_2$, because excluding $q_2$ only slightly changed the result (Fig. \ref{fig:si_xai_wat}).
Similarly to the case in vacuum, CV${}_{61}$, CV${}_{58}$, and CV${}_{54}$, corresponding to $\phi$, $\phi$, and $\theta$, respectively, are found to be the three major CVs.
Apart from these three CVs, CV${}_{125}$, the electrostatic potential from the water on H${}_{18}$ ({\vele}) (Fig. \ref{fig:molecule}), shows up as a key feature from the solvent.
In addition, CV${}_{105}$ and CV${}_{119}$, which are the electrostatic potential on H${}_{8}$ and C${}_{15}$, respectively, show notable contribution to the RC.
Same trend is found in the SHAP result though the relative balance is somewhat changed.
The scatter plots of {\vele} and $\phi$ or $\theta$, given in Figs. \ref{fig:corr_cvs_pb}(a) and (b), respectively, do not show a clear correlation between the changes of CVs and {\pb} or any apparent separatrix. On the other hand, the plot as a function of the three variables (Fig. \ref{fig:corr_cvs_pb}(c)) shows that there is a weak correlation between {\vele} and {\pb} near the separatrix in the ($\phi$, $\theta$) space.
Thus the solvation coordinate {\vele} is contributing to the RC in a nontrivial manner.
Interestingly, the importance of solvent effect to H${}_{18}$ have also been indicated by Ma and Dinner\cite{Ma.2005} for the C${}_{\mathrm{7eq}}$ $\leftrightharpoons$ $\alpha_{\mathrm{R}}$ transition through the torque coordinate as mentioned above.
\rev{The structures about the TS for $q_1$ is also summarized in Figure \ref{fig:si_structre_ts_q1}. The figure shows that the position of H${}_{18}$ strongly depends on $\psi$, which ranges between $-123^{\circ} < \psi < -50^{\circ}$. A water molecule is frequently placed near H${}_{18}$ but without clear orientation. This confirms that the solvent contributes to the reaction coordinate in a collective manner.}

Finally, the RCs in water are directly compared in Fig. \ref{fig:si_corr_wat}.
Compared to the vacuum results (Fig. \ref{fig:si_corr_vac}), the deviation from the diagonal line is slightly larger even at $q_i \sim 0$ (i.e., near TS).
Nevertheless, the overall correlation between the RCs are very high and found to be above 0.96 except for $q_2$.
Furthermore, the correlation between $q_2$, which was indicated to be slightly overfitted, and other CVs is still above 0.94.
Thus, the current hyperparameter tuning framework successfully obtained the RC for the alanine dipeptide isomerization in water, where the DNN models can differ but the important features remain very similar.

%
%
%

%
%

%
%

%
%
\subsection{On the efficiency of hyperparameter tuning}

Since the hyperparameter space defined in the current set up is large, i.e., the number of grids in the nodes and regularization parameter per node are 50 and 20, respectively, 150 cycle in the Bayesian optimization cannot explore the whole space.
It can thus be anticipated that the diversity in the model is a consequence of insufficient sampling despite each model shows a similar performance.

\begin{figure}
\includegraphics[width=8.5cm]{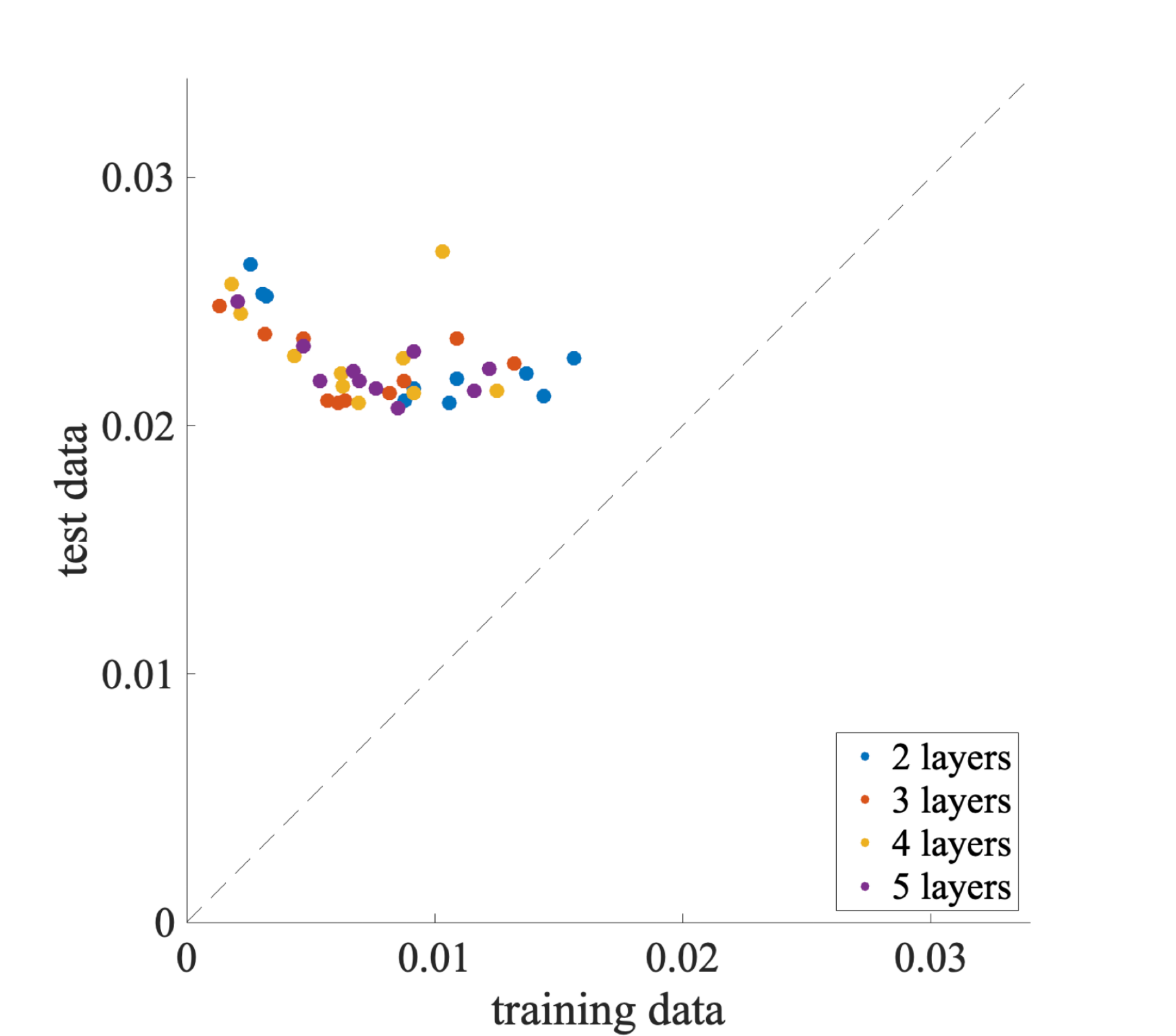}
\caption{Scatter plot for RMSEs between the predicted and reference {\pb} for the training and test data sets where the number of layers are fixed during hyperparameter tuning. Blue, red, yellow, and purple points are from the DNN model with {\nlayer} fixed to 2, 3, 4, and 5, respectively.}
\label{fig:rmse_extensive}
\end{figure}

\begin{figure}
\includegraphics[width=14cm]{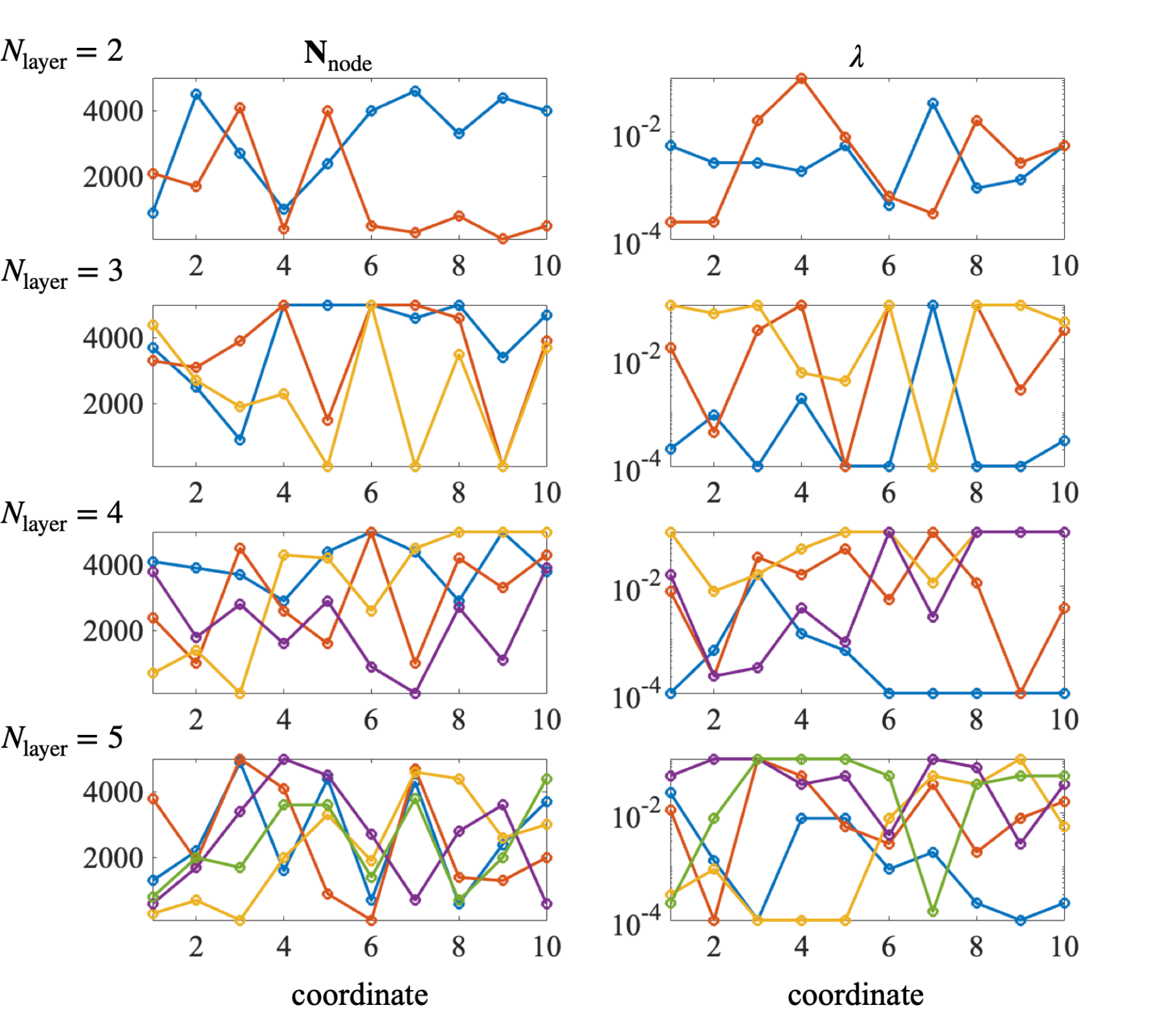}
\caption{Optimized hyperparameters for RCs in water with fixed number of layers during hyperparameter tuning.
Blue, red, yellow, purple, and green lines indicate the number of nodes and regularization parameters for layers 1, 2, 3, 4, and 5, respectively.
}
\label{fig:si_hp_optmized_extensive}
\end{figure}

To this end, we more intensely explored the hyperparameter space for the RCs in water using fixed number of layers and 500 Bayesian optimization cycles.
The optimized hyperparameters for different number of fixed layers are summarized in Fig. \ref{fig:si_hp_optmized_extensive}, and the RMSEs for these models are given in Fig. \ref{fig:rmse_extensive}. 
Figure \ref{fig:si_hp_optmized_extensive} shows that even when the exploration space is restricted and more carefully investigated, the hyperparameters do not converge to a unique value, which confirms that the hyperparameter space is indeed multimodal.
The RMSEs for different {\nlayer} in Fig. \ref{fig:rmse_extensive} indicate that the RMSE for the training data tend to improve when the number of layers are increased, whereas the predictability (i.e. RMSE for the test data) do not improve.
This may be the reason that {\nlayer} was somewhat smaller in the case in water compared to that in vacuum.
Importantly, the RMSEs in Fig. \ref{fig:rmse_extensive} do not show apparent improvement compared to that with less number of Bayesian trials (Fig. \ref{fig:rmse_train_test}).
These results show that the current Bayesian optimization of 150 cycle was effective in obtaining the optimal DNN model for the RC from the large number of hyperparameter candidates.
The result also highlights that balance between accuracy (RMSE for training data) and predictability (RMSE for test data) is delicate.
%
%

%
%
\section{Summary} \label{sec:summary}

Machine learning approaches have become a powerful tool in determining the RCs from many CV candidates for the reactions in complex environments.
DNN is widely used for its effectiveness in taking account of the non-linear contribution of the CVs to the RC, and XAI tools serves as a complement to understand the features that characterize RC which is otherwise hidden in the complex DNN structure.
On the other hand, the structure of DNN model can be highly flexible, and the hyperparameters that determine the structure are often determined in a non-trivial manner and remains to be a tedious task.
Here, we developed the hyperparameter tuning approach by utilizing the Bayesian optimization method to determine the DNN models for the RC.
The DNN model was optimized so as to obtain a RC that can predict the changes of committors from the reactant to product via cross-entropy minimization.

The current approach was first applied to analyze the isomerization of alanine dipeptide in vacuum.
The RC was successfully obtained from 10 different initial conditions, where the RMSE of {\pb} between the prediction from the RC and the actual data was about 0.005.
The correlations between the 10 RCs were also very high ($>$0.99).
On the other hand, the structure of the optimized DNN model, i.e., the hyperparameters, varied notably between the RCs.
By applying the LIME and SHAP methods, all RCs \rev{were} found to have the same key features, i.e., $\phi$ and $\theta$.
Thus, despite the apparent differences in the DNN models, all RCs share common physical mechanism for the reaction in vacuum with similar accuracy.

This approach was further applied to the isomerization in water.
The RC for the reaction in solution was successfully obtained in most cases (9 out of 10) where the RMSEs between the predicted and calculated {\pb} for the test data were about 0.02.
In one case ($q_2$) we found slight sign of overfitting, where the committor probability distribution and RMSEs for training and test data sets showed a discrepancy of $\sim$0.02 in {\pb}.
Similarly to the case in vacuum, the hyperparameters were found to vary notably, but the successfully optimized RCs showed similar {\pb}-predictability and high correlation ($>$ 0.96).
By analyzing the RCs using the XAI methods, the RCs were found to have common key features, 
i.e., $\phi$, $\theta$, and the electrostatic potential from the water on H${}_{18}$.
We note that this solvent contribution to the hydrogen was also found in the torque coordinate proposed previously for the different transition path of the same system\cite{Ma.2005}.
The current RC was able to describe this from a rather simple set of CV candidates while the complex non-linear contribution is taken into account through the DNN model structure.
\rev{We note that choosing the good CV candidates are important for improving the quality of the RC, such as utilizing the graph neural networks.}

The hyperparameter tuning framework was shown to be applicable to explore the RC for reactions in different systems straightforwardly.
It is noted that here the RCs were optimized using slightly different data set, i.e., the same data was partitioned into slightly different training, validation, and test groups due to different random seed.
Even when the same data set was used, the optimization converged to different optimal when different initial hyperparameters were used (not shown).
Thus, the hyperparameter space of the DNN model for the RC is likely multimodal, and optimization of the hyperparamters can converge to different models with similarly accurate RC depending on initial conditions.
On the other hand, the application of the XAI tools to these different DNN models indicates that suitably optimized DNN models share the same features, thus the same mechanism can be extracted from the apparently different DNN models when the hyperparemters are optimized adequately.

\section*{Supplementary material}
\begin{figure*}
\includegraphics[width=14cm]{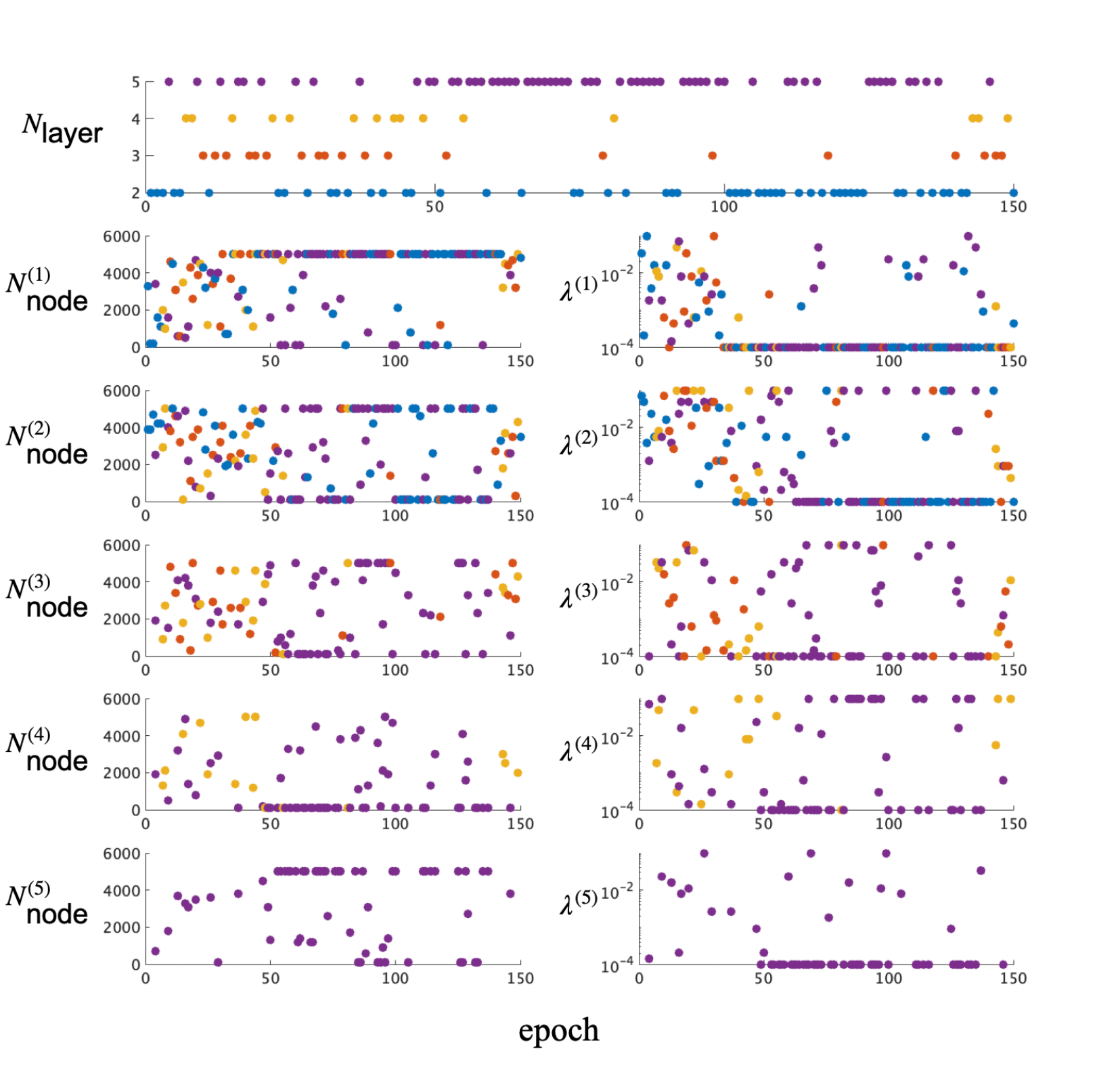}
\caption{History of the hyperparameters explored during the hyperparameter tuning for $q_1$ in vacuum. Blue, red, yellow, and purple indicate the epochs with {\nlayer} of 2, 3, 4, and 5, respectively.}
\label{fig:si_hp_history_vac}
\end{figure*}

\begin{figure*}
\includegraphics[width=14cm]{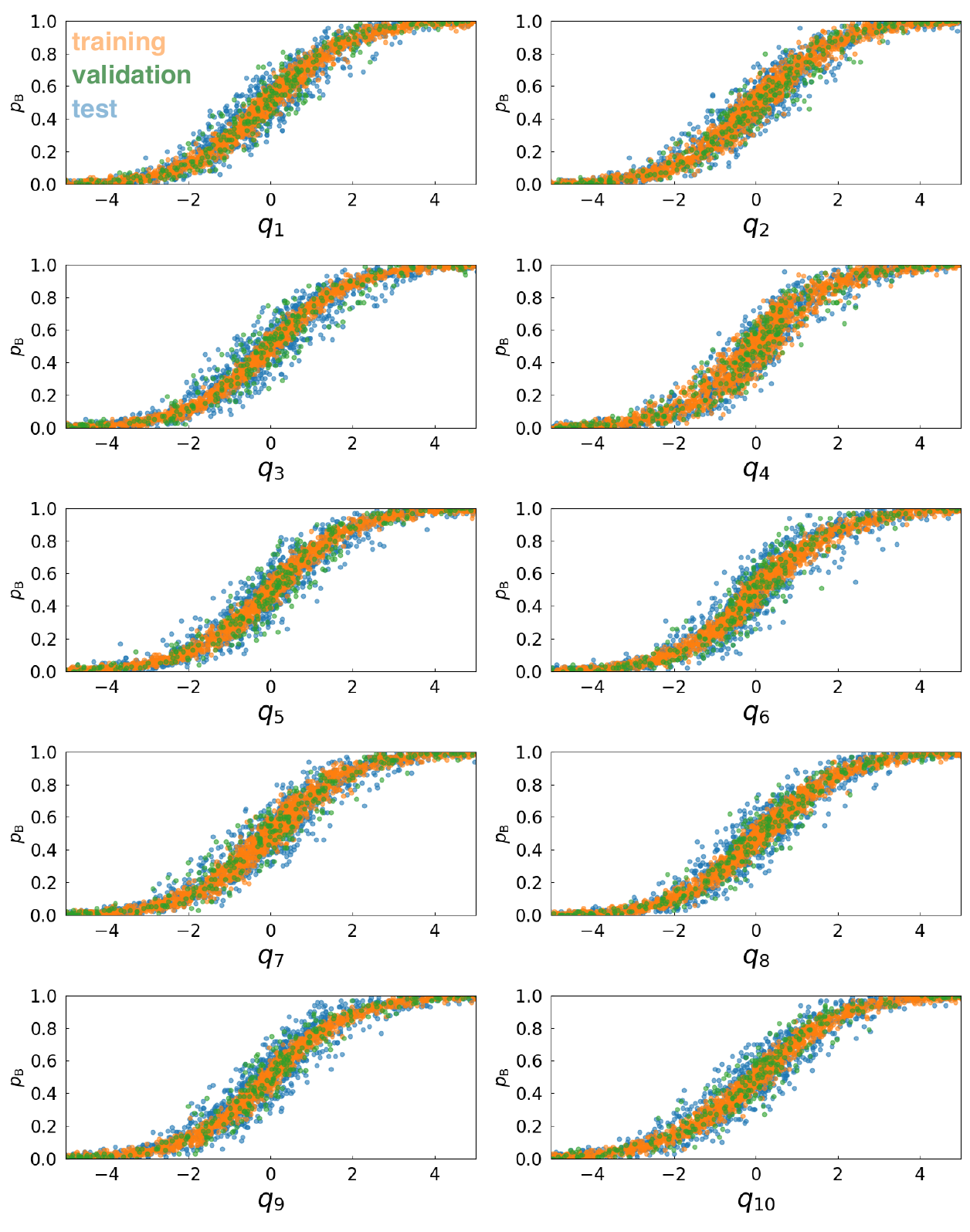}
\caption{Scatter plots of the optimized coordinates ($q_1$ to $q_{10}$) and committors ($p_\mathrm{B}$) in vacuum. Orange, green, and blue in (a) and (b) denote the results from the training, validation, and test data sets, respectively.}
\label{fig:si_sigmoid_ala_vac}
\end{figure*}

\begin{figure*}
\includegraphics[width=14cm]{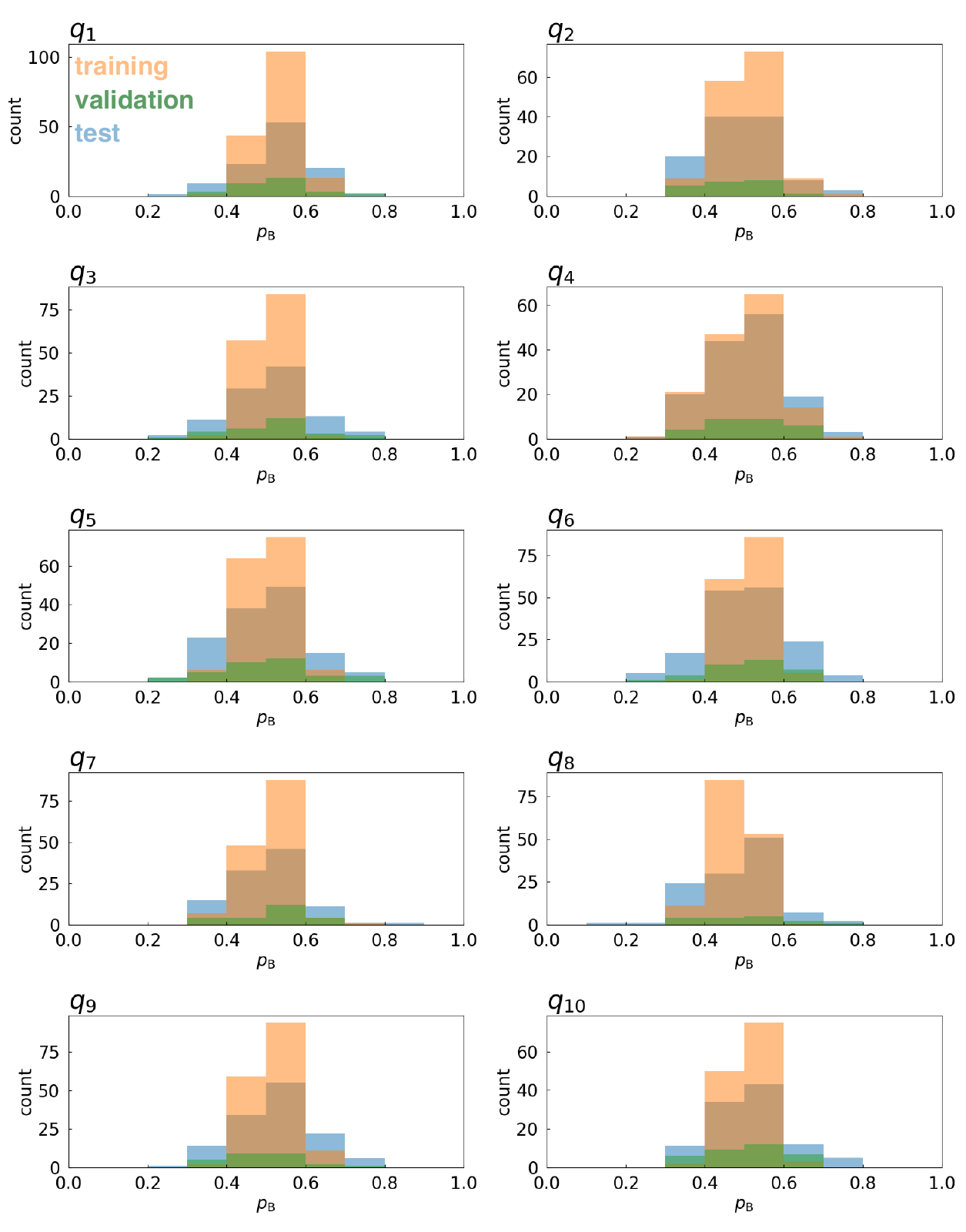}
\caption{Histogram of the committors for the data points within $-0.2 < q < 0.2$ for $q_1$ to $q_{10}$ in vacuum. Orange, green, and blue in (a) and (b) denote the results from the training, validation, and test data sets, respectively.}
\label{fig:si_hist_ala_vac}
\end{figure*}

\begin{figure*}
\includegraphics[width=14cm]{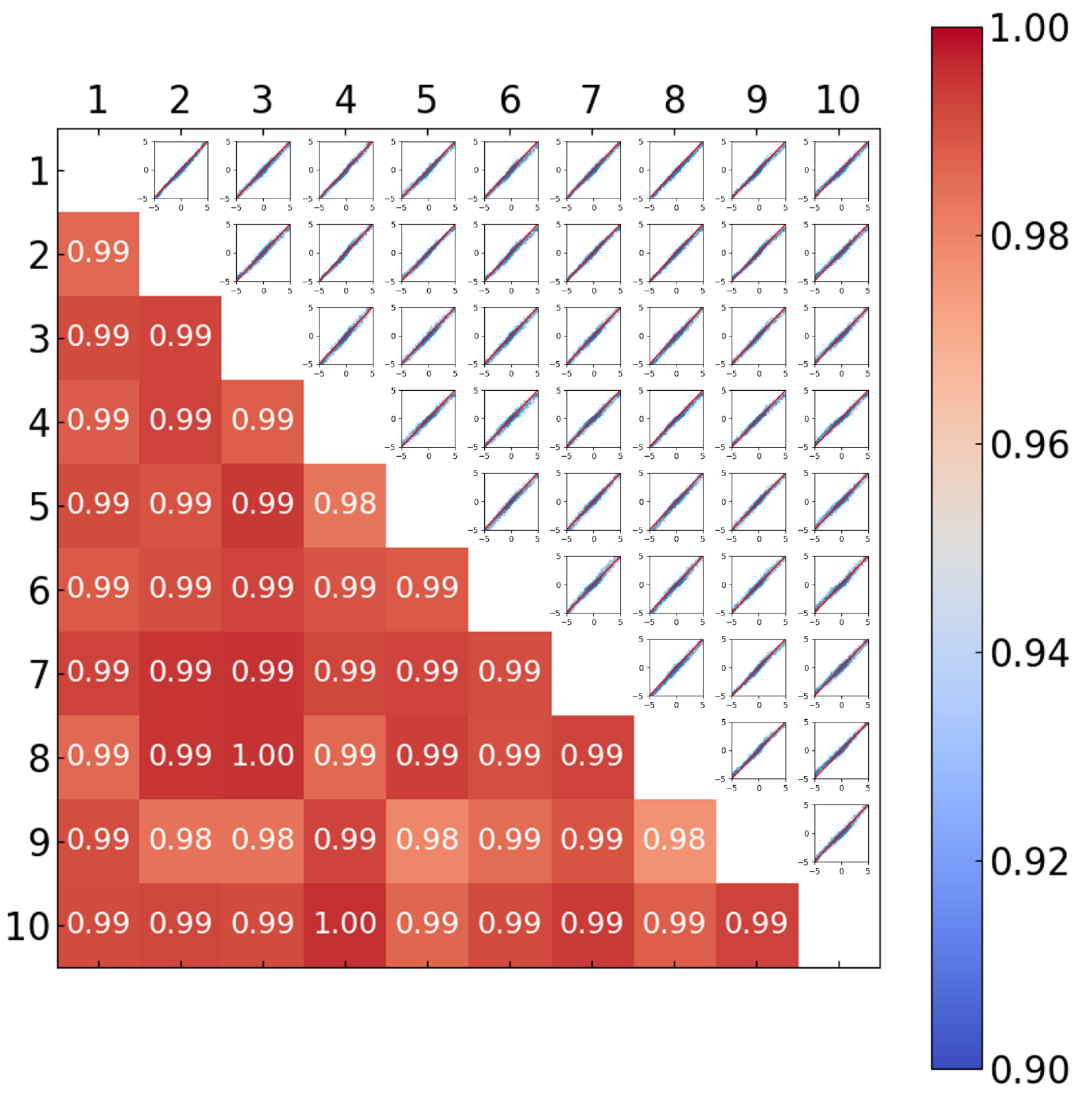}
\caption{Scatter plots (top-right) and correlation coefficient (bottom-left) showing the correlation between the RCs from different optimized DNN models in vacuum.
$q_{i}$ denote the optimized coordinate from the $i$-th DNN model.
Note that all data points are included in these plots.
}
\label{fig:si_corr_vac}
\end{figure*}

\begin{figure*}
\includegraphics[width=14cm]{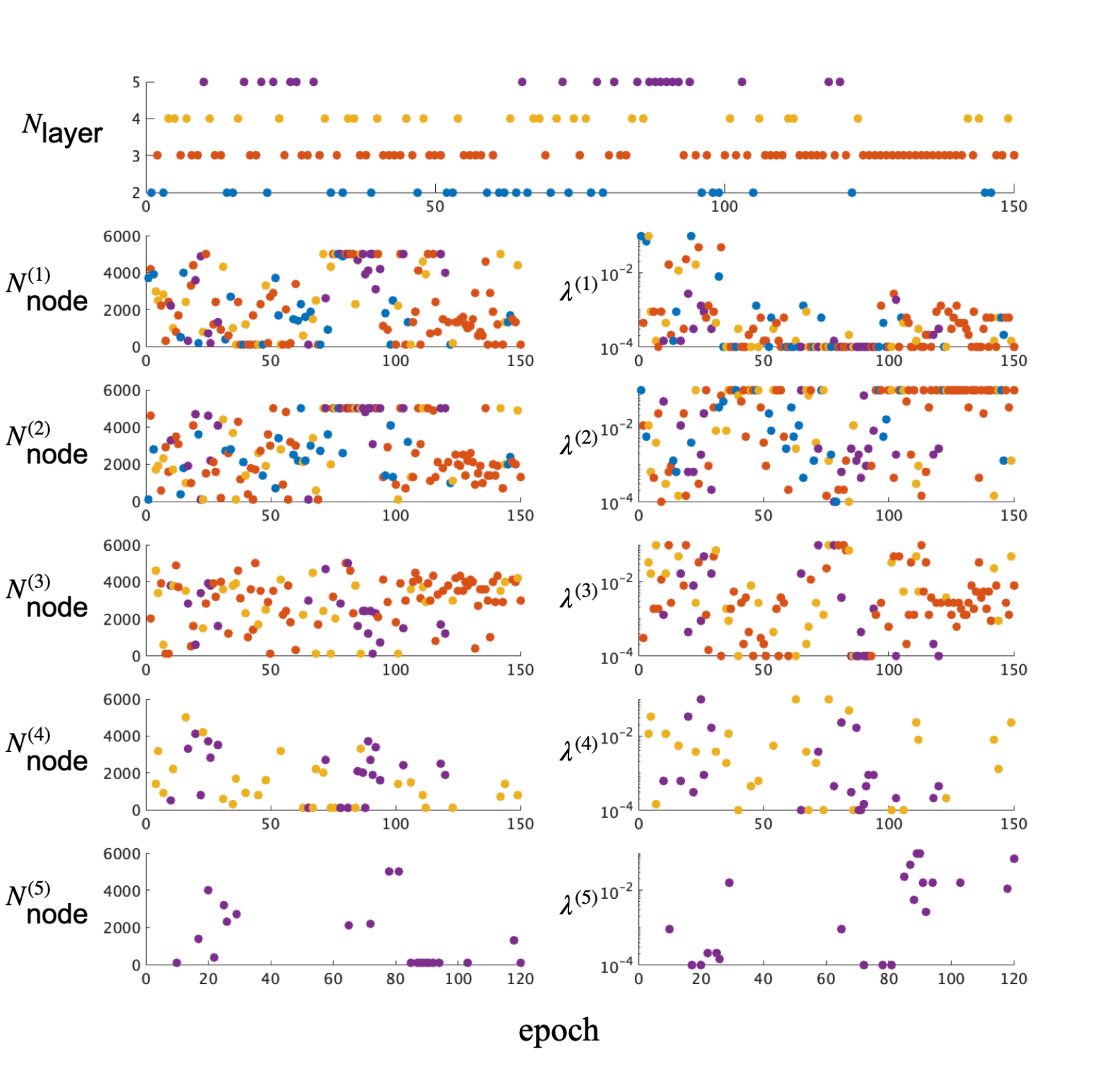}
\caption{History of the hyperparameters explored during the hyperparameter tuning for $q_1$ in water. Blue, red, yellow, and purple indicate the epochs with {\nlayer} of 2, 3, 4, and 5, respectively.}
\label{fig:si_hp_history_wat}
\end{figure*}

\begin{figure*}
\includegraphics[width=14cm]{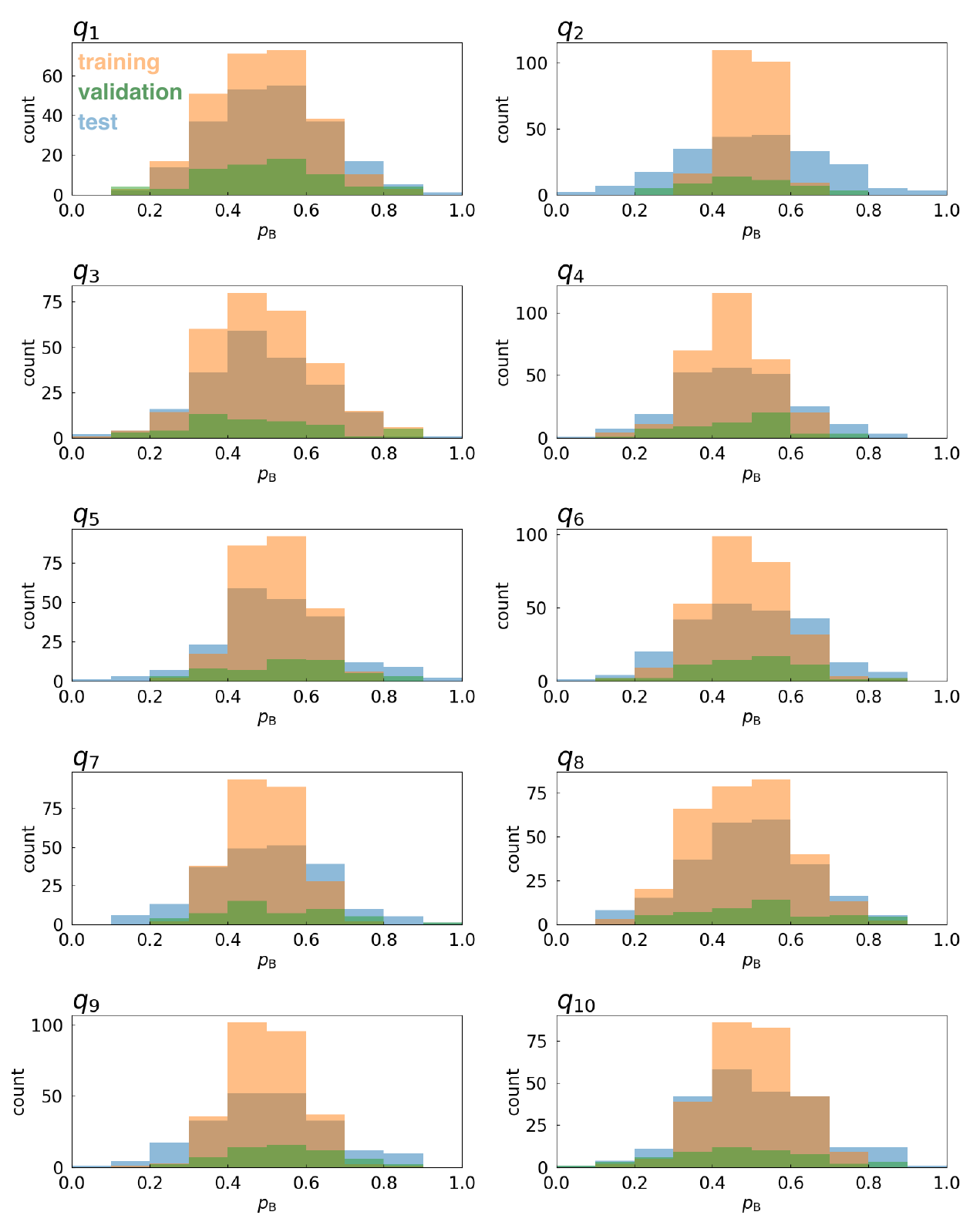}
\caption{Histogram of the committors for the data points within $-0.2 < q < 0.2$ for $q_1$ to $q_{10}$ in water. Orange, green, and blue in (a) and (b) denote the results from the training, validation, and test data sets, respectively.}
\label{fig:si_hist_ala_wat}
\end{figure*}

\begin{figure*}
\includegraphics[width=14cm]{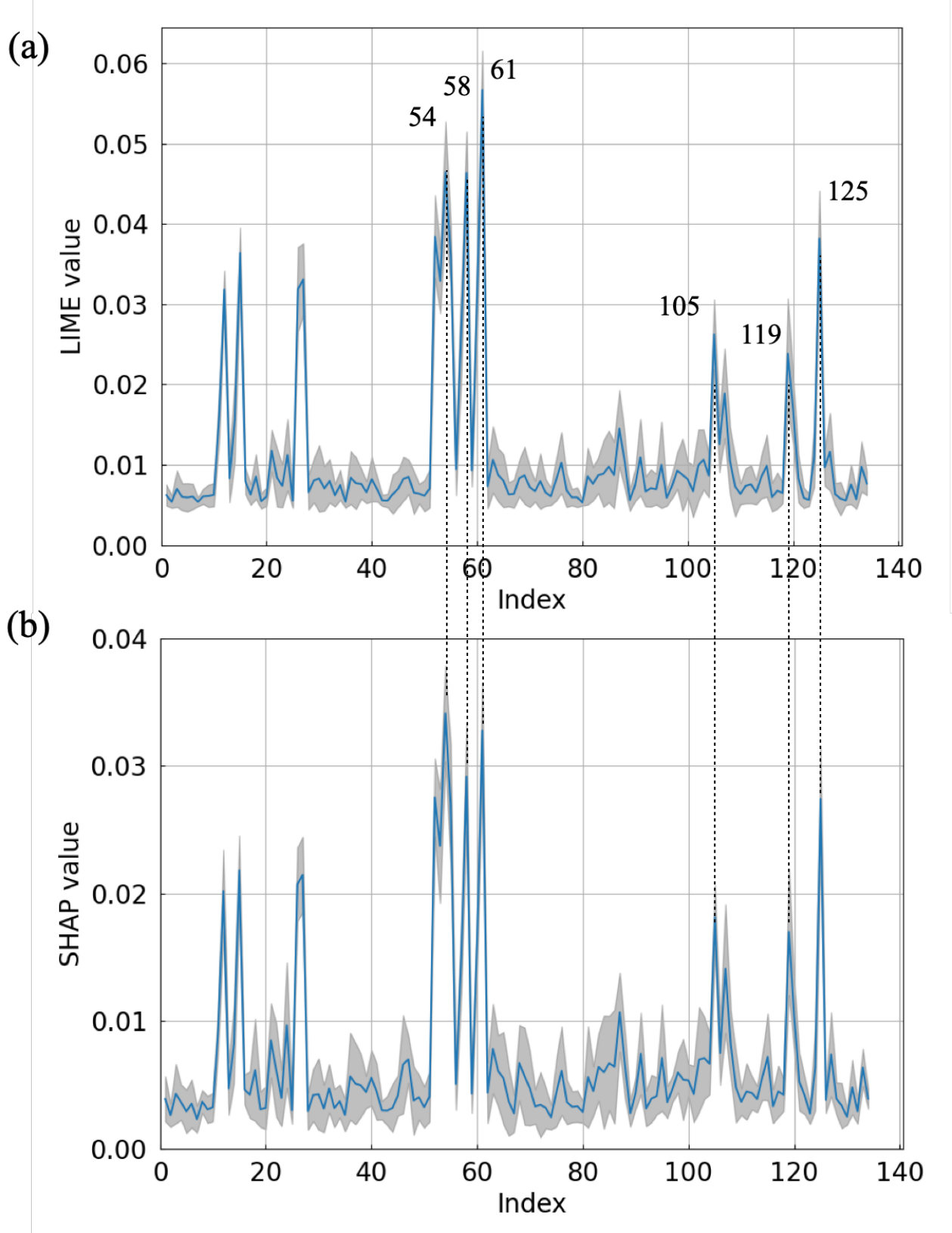}
\caption{Contributions of CVs to RCs in water excluding $q_2$ extracted using (a) LIME and (b) SHAP in absolute values. Blue lines and gray shades denote the average and variance calculated from the 9 RCs.}
\label{fig:si_xai_wat}
\end{figure*}

\begin{figure*}
\includegraphics[width=14cm]{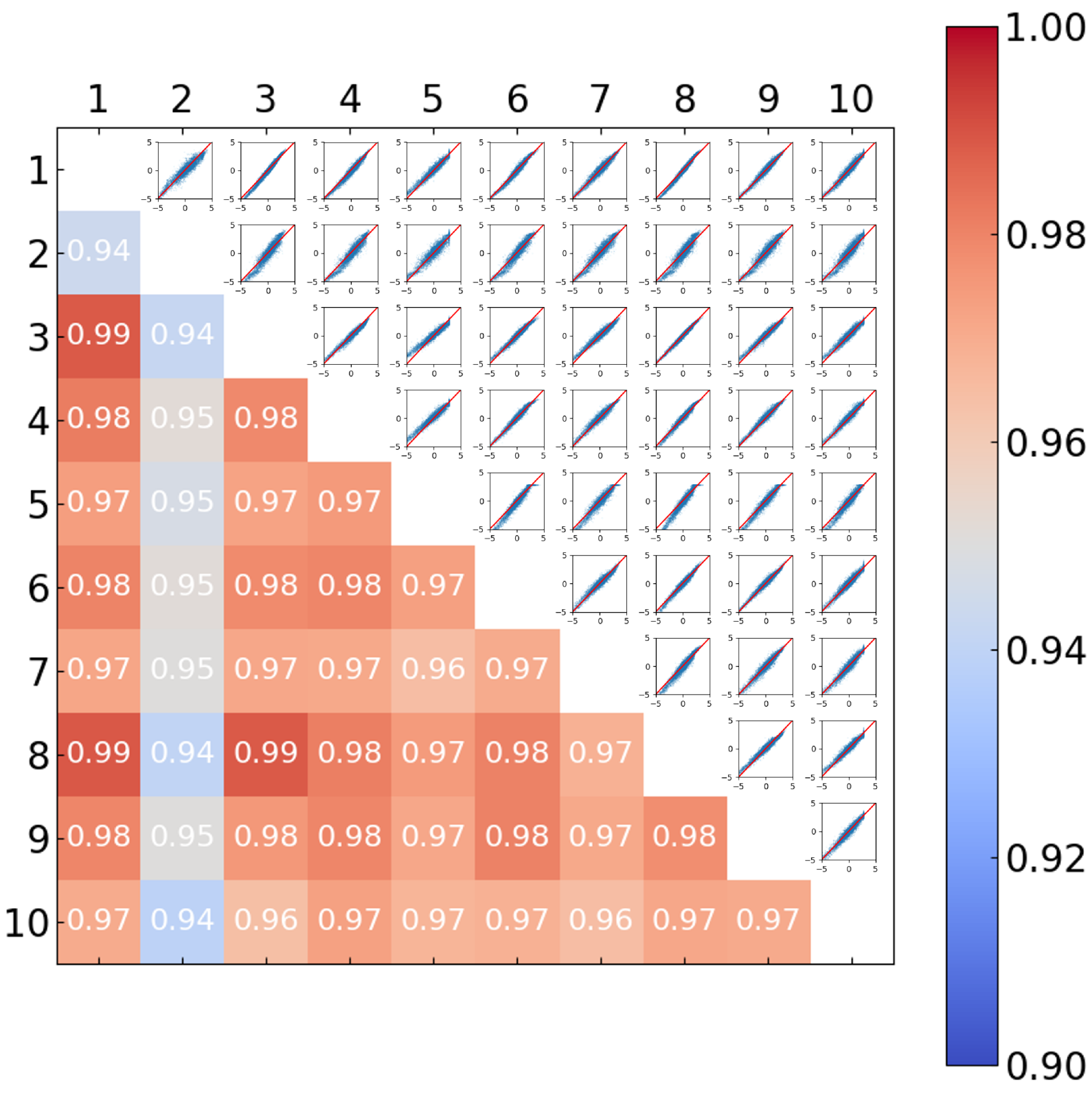}
\caption{Scatter plots (top-right) and correlation coefficient (bottom-left) showing the correlation between the RCs from different optimized DNN models in water.
$q_{i}$ denote the optimized coordinate from the $i$-th DNN model.
Note that all data points are included in these plots.
}
\label{fig:si_corr_wat}
\end{figure*}

\begin{figure}
\includegraphics[width=8.5cm]{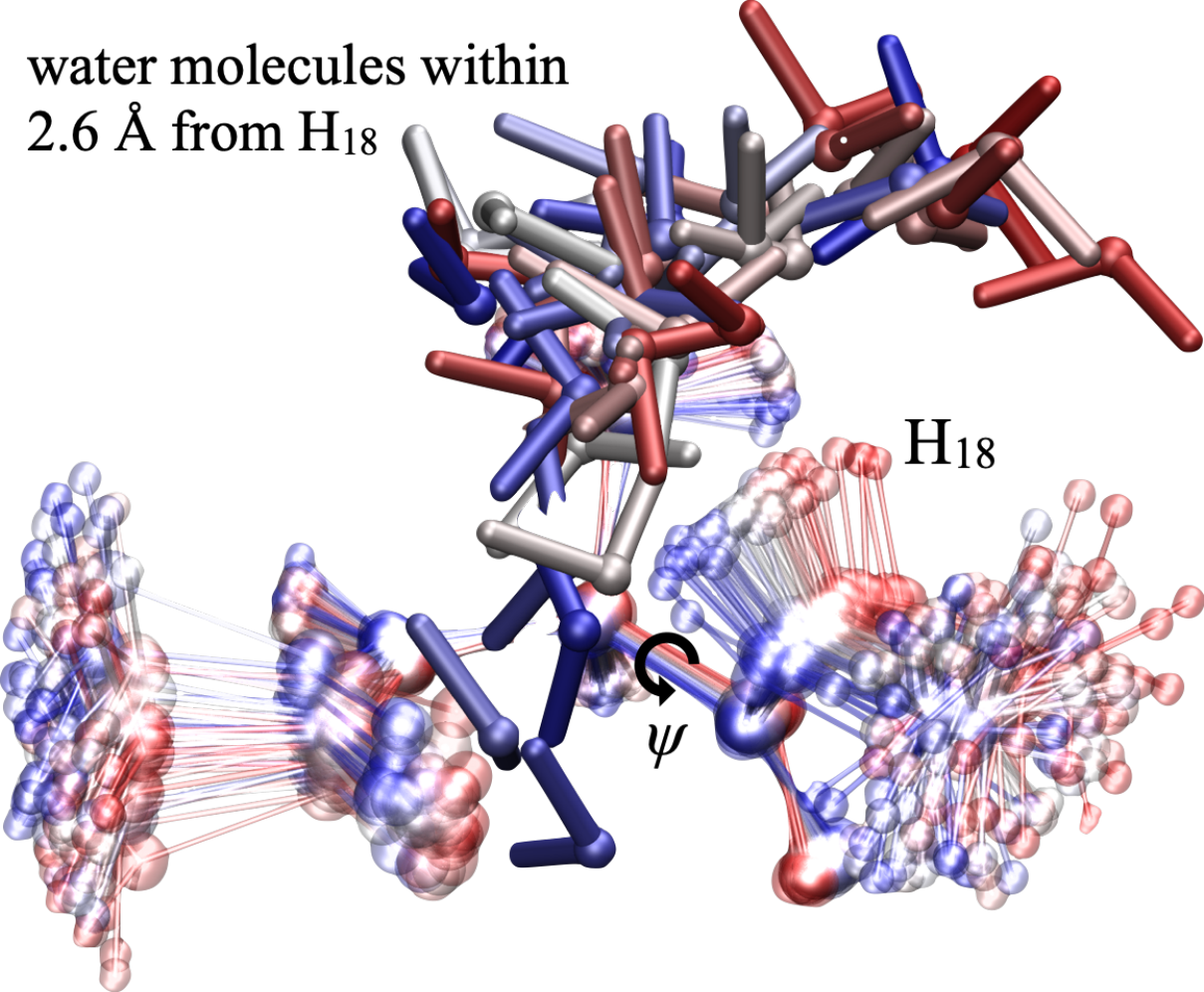}
\caption{\rev{Structures within $-0.2 < q_1 < 0.2$ sorted based on $\psi$ (here defined as the dihedral angle of N-CA-C-N) and aligned against the backbone atoms of alanine. RWB color range indicate change of $\psi$ in the structures from -123${}^{\circ}$ to -50${}^{\circ}$. Water molecules within 2.6 {\AA} from H${}_{18}$, which is within the first solvation shell, is shown in licorice, whereas the alanine dipeptide is given in transparent color.}
}
\label{fig:si_structre_ts_q1}
\end{figure}

\rev{Full list of collective variables in vacuum and in water (Table \ref{table:si_cvs}), history of hyperparameters searched in vacuum (Fig. \ref{fig:si_hp_history_vac}), results for all optimized coordinates in vacuum (Figs. \ref{fig:si_sigmoid_ala_vac}, \ref{fig:si_hist_ala_vac}, and \ref{fig:si_corr_vac}), history of hyperparameters searched in water (Fig. \ref{fig:si_hp_history_wat}), results for all optimized coordinates in water (Figs. \ref{fig:si_sigmoid_ala_wat}, \ref{fig:si_hist_ala_wat}, and \ref{fig:si_corr_wat}), LIME and SHAP analyses without $q_2$ in water (Fig. \ref{fig:si_xai_wat}), history of hyperparameters searched with fixed number of layers (Fig. \ref{fig:si_hp_optmized_extensive}), and the structures at about the TS in water (Fig. \ref{fig:si_structre_ts_q1}).}

%
%
\begin{acknowledgments} 
This work was supported by Grant-in-Aid for Scientific Research (JP22H02035, JP23K23303, JP23KK0254, JP24K21756, JP22H02595, JP22K03550, JP23H02622, JP23K23858, JP23K27313, JP24H01719) from JSPS.
The calculations were partially carried out at the Research Center for Computational Sciences in Okazaki (23-IMS-C111, \rev{24-IMS-C051}, 24-IMS-C105, and 24-IMS-C198) and \rev{using MCRP-M at the Center for Computational Sciences, University of Tsukuba}.
T.M. also acknowledges the support from Pan-Omics Data-Driven Research Innovation Center, Kyushu University.
\end{acknowledgments}

\bibliography{manuscript_HPtuning}

\section*{AUTHOR DECLARATIONS}

\section*{Conflict of Interest}
The authors have no conflicts of interest to disclose.

\section*{Data availability}
The data that support the findings of this study are openly available in Zenodo at https://doi.org/10.5281/zenodo.14378447.

\clearpage
\widetext
%
%
\beginsupplement

\noindent{\bf\Large Supplementary Material}
\vspace{2mm}
\begin{center}
\textbf{\Large \titlename}

\vspace{5mm}

{Kyohei Kawashima$^{1}$, Takumi Sato$^{2}$, Kei-ichi Okazaki$^{3,4,a)}$, Kang Kim$^{5,b)}$, Nobuyuki Matubayasi$^{5,c)}$, and Toshifumi Mori$^{1,2,d)}$}
\end{center}

\vspace{5mm}
\noindent
\textit{%
$^{1)}$Institute for Materials Chemistry and Engineering, Kyushu University, Kasuga, Fukuoka 816-8580, Japan}\\
\textit{
$^{2)}$Department of Interdisciplinary Engineering Sciences, Interdisciplinary Graduate School of Engineering Sciences, Kyushu University, Kasuga, Fukuoka 816-8580, Japan}\\
\textit{
$^{3)}$Research Center for Computational Science, Institute for Molecular Science, Okazaki, Aichi 444-8585, Japan} \\
\textit{
$^{4)}$Graduate Institute for Advanced Studies, SOKENDAI, Okazaki,
Aichi 444-8585, Japan} \\
\textit{
$^{5)}$Division of Chemical Engineering, Department of Materials Engineering Science, Graduate School of Engineering Science, Osaka University, Toyonaka, Osaka 560-8531, Japan}

\vspace{5mm}

\noindent
\small{$^{a)}$Electronic mail: keokazaki@ims.ac.jp \\
$^{b)}$Electronic mail: kk@cheng.es.osaka-u.ac.jp \\
$^{c)}$Electronic mail: nobuyuki@cheng.es.osaka-u.ac.jp \\
$^{d)}$Electronic mail: toshi\underline{ }mori@cm.kyushu-u.ac.jp
}

\begin{table*}
\caption{Definitions of collective variables used for the reactions in vacuum and water.}
\centering
      \begin{tabular}{c|lll}
      \multicolumn{4}{l}{dihedrals (CV 1-45: cosine, 46-90: sine)} \\ \hline
      index & \multicolumn{3}{c}{atom number} \\ \hline
      $01-03$ &  1 -  2 -  5 -  6 &  1 -  2 -  5 -  7  &  3 -  2 -  5 -  6 \\
      $04-06$ &  3 -  2 -  5 -  7 &  4 -  2 -  5 -  6  &  4 -  2 -  5 -  7 \\
      $07-09$ &  2 -  5 -  7 -  8 &  2 -  5 -  7 -  9  &  6 -  5 -  7 -  8 \\
      $10-12$ &  6 -  5 -  7 -  9 &  5 -  7 -  9 - 10  &  5 -  7 -  9 - 11 \\
      $13-15$ &  5 -  7 -  9 - 15 &  8 -  7 -  9 - 10  &  8 -  7 -  9 - 11 \\
      $16-18$ &  8 -  7 -  9 - 15 &  7 -  9 - 11 - 12  &  7 -  9 - 11 - 13 \\
      $19-21$ &  7 -  9 - 11 - 14 & 10 -  9 - 11 - 12  & 10 -  9 - 11 - 13 \\
      $22-24$ & 10 -  9 - 11 - 14 & 15 -  9 - 11 - 12  & 15 -  9 - 11 - 13 \\
      $25-27$ & 15 -  9 - 11 - 14 &  7 -  9 - 15 - 16  &  7 -  9 - 15 - 17 \\
      $28-30$ & 10 -  9 - 15 - 16 & 10 -  9 - 15 - 17  & 11 -  9 - 15 - 16 \\
      $31-33$ & 11 -  9 - 15 - 17 &  9 - 15 - 17 - 18  &  9 - 15 - 17 - 19 \\
      $34-36$ & 16 - 15 - 17 - 18 & 16 - 15 - 17 - 19  & 15 - 17 - 19 - 20 \\
      $37-39$ & 15 - 17 - 19 - 21 & 15 - 17 - 19 - 22  & 18 - 17 - 19 - 20 \\
      $40-42$ & 18 - 17 - 19 - 21 & 18 - 17 - 19 - 22  & 2 -  7 -  5 -  6 \\
      $43-45$ &  5 -  9 -  7 -  8 &  9 - 17 - 15 - 16  & 15 - 19 - 17 - 18 \\
      \hline \multicolumn{4}{c}{} \\
      \multicolumn{4}{l}{electrostatic (V${}_\mathrm{ele}$) \& van der Waals (V${}_\mathrm{vdW}$) potentials} \\ \hline
      index & \multicolumn{3}{c}{atom number} \\ \hline
      $91-93$ & V${}_\mathrm{ele}(1)$ & V${}_\mathrm{vdW}(1)$ & V${}_\mathrm{ele}(2)$ \\
	  $94-96$ &  V${}_\mathrm{vdW}(2)$ & V${}_\mathrm{ele}(3)$ & V${}_\mathrm{vdW}(3)$ \\
      $97-99$ & V${}_\mathrm{ele}(4)$ & V${}_\mathrm{vdW}(4)$ & V${}_\mathrm{ele}(5)$ \\
	  $100-102$ &  V${}_\mathrm{vdW}(5)$ & V${}_\mathrm{ele}(6)$ & V${}_\mathrm{vdW}(6)$ \\
      $103-105$ & V${}_\mathrm{ele}(7)$ & V${}_\mathrm{vdW}(7)$ & V${}_\mathrm{ele}(8)$ \\
	  $106-108$ &  V${}_\mathrm{vdW}(8)$ & V${}_\mathrm{ele}(9)$ & V${}_\mathrm{vdW}(9)$ \\
      $109-111$ & V${}_\mathrm{ele}(10)$ & V${}_\mathrm{vdW}(10)$ & V${}_\mathrm{ele}(11)$ \\
	  $112-114$ &  V${}_\mathrm{vdW}(11)$ & V${}_\mathrm{ele}(12)$ & V${}_\mathrm{vdW}(12)$ \\
      $115-117$ & V${}_\mathrm{ele}(13)$ & V${}_\mathrm{vdW}(13)$ & V${}_\mathrm{ele}(14)$ \\
	  $118-120$ &  V${}_\mathrm{vdW}(14)$ & V${}_\mathrm{ele}(15)$ & V${}_\mathrm{vdW}(15)$ \\
      $121-123$ & V${}_\mathrm{ele}(16)$ & V${}_\mathrm{vdW}(16)$ & V${}_\mathrm{ele}(17)$ \\
	  $124-126$ &  V${}_\mathrm{vdW}(17)$ & V${}_\mathrm{ele}(18)$ & V${}_\mathrm{vdW}(18)$ \\
      $127-129$ & V${}_\mathrm{ele}(19)$ & V${}_\mathrm{vdW}(19)$ & V${}_\mathrm{ele}(20)$ \\
	  $130-132$ &  V${}_\mathrm{vdW}(20)$ & V${}_\mathrm{ele}(21)$ & V${}_\mathrm{vdW}(21)$ \\
      $133-134$ & V${}_\mathrm{ele}(22)$ & V${}_\mathrm{vdW}(22)$ \\
      \end{tabular}
\label{table:si_cvs}
\end{table*}

%
%

%
%

%
%

%
%

%
%

%
%

%
%

%
%

%
%

%
%

%
%


\end{document}